%% file: mainhpca.tex
\documentclass[camera,letterpaper,nomarginnotes,nonarrowgutter]{jpaper}
\usepackage{cite}
\usepackage{amsmath,amssymb,amsfonts}
\usepackage{algorithmic}
\usepackage{graphicx}
\usepackage{textcomp}
\usepackage{xcolor}
\usepackage{comment}
\usepackage{pifont}
\usepackage{siunitx}
\usepackage{setspace} %
\usepackage{flushend}
\usepackage[normalem]{ulem}

\usepackage{tikz}
\usetikzlibrary{calc}
\newcommand*\circled[1]{\tikz[baseline=(char.base)]{
    \node[shape=circle, draw, fill, inner sep=0.05pt,] (char) {\vphantom{WAH1g}\textcolor{white}{#1}};}}

\def\BibTeX{{\rm B\kern-.05em{\sc i\kern-.025em b}\kern-.08em
    T\kern-.1667em\lower.7ex\hbox{E}\kern-.125emX}}

\usepackage{mathptmx} %
\usepackage[italic]{mathastext}

\usepackage[dvipsnames]{xcolor}
\usepackage{fancyhdr}
\usepackage[normalem]{ulem}
\usepackage{multirow}
\usepackage{enumitem}
\usepackage{algorithm}
\usepackage{listings}
\usepackage{amsmath}
\usepackage{arydshln} %
\usepackage{xspace}
\usepackage{soul} %
\usepackage[bookmarks=true,breaklinks=true,hidelinks]{hyperref}
\usepackage{amssymb}%
\usepackage{pifont}%
\usepackage{setspace}
\usepackage{footmisc}
\usepackage{balance}
\usepackage{hhline} %
\usepackage[all]{nowidow}

\usepackage[compact]{titlesec}
\usepackage{footnote}

\definecolor{amethyst}{rgb}{0.6, 0.4, 0.8}
\definecolor{amber}{rgb}{1.0, 0.49, 0.0}
\definecolor{awesome}{rgb}{1.0, 0.13, 0.32}
\definecolor{dollarbill}{rgb}{0.52,0.73,0.4}
\definecolor{moegi}{rgb}{0.357, 0.537, 0.188}
\definecolor{burgundy}{rgb}{0.5, 0.0, 0.13}
\definecolor{ballblue}{rgb}{0.13, 0.67, 0.8}
\definecolor{ups-truck}{rgb}{0.53, 0.28, 0.21}
\definecolor{airforceblue}{rgb}{0.36, 0.54, 0.66}
\definecolor{cadmiumgreen}{rgb}{0.0, 0.42, 0.24}
\definecolor{darkcyan}{rgb}{0.0, 0.55, 0.55}
\definecolor{caribbeangreen}{rgb}{0.0, 0.8, 0.6}
\definecolor{flamingopink}{rgb}{0.99, 0.56, 0.67}
\definecolor{jazzberryjam}{rgb}{0.65, 0.04, 0.37}
\definecolor{mediumpersianblue}{rgb}{0.0, 0.4, 0.65}
\definecolor{coolblack}{rgb}{0.0, 0.18, 0.39}
\definecolor{bleudefrance}{rgb}{0.19, 0.55, 0.91}
\definecolor{ao}{rgb}{0.0, 0.0, 1.0}
\definecolor{babyblueeyes}{rgb}{0.63, 0.79, 0.95}
\definecolor{darkblue}{rgb}{0.0, 0.0, 0.55}
\definecolor{lava}{rgb}{0.81, 0.06, 0.13}
\definecolor{gfored}{rgb}{0.580, 0.050, 0.211}

\newif\ifcommentoff
\commentofftrue
\ifcommentoff
\newcommand{\agycomment}[1]{}
\newcommand{\omc}[1]{}

\newcommand{\nisf}[1]{\textcolor{black}{#1}}
\else

\fi

\newif\ifcamready
\camreadyfalse

\ifcamready

\newcommand{\revision}[1]{\textcolor{black}{#1}}

\newcommand{\om}[1]{\textcolor{black}{#1}}
\newcommand{\omc}[1]{}
\newcommand{\nisd}[1]{}
\newcommand{\nise}[1]{\textcolor{black}{#1}}
\newcommand{\ctodo}[2][]{}

\newcommand{\agycomment}[1]{}

\else

\newcommand{\revision}[1]{\textcolor{black}{#1}}

\newcommand{\om}[1]{\textcolor{black}{#1}}

\newcommand{\nisd}[1]{\textcolor{cadmiumgreen}{[\textbf{Nisa:} #1]}}
\newcommand{\nise}[1]{\textcolor{black}{#1}}
\newcommand{\ctodo}[2][fix]{\textbf{\emph{\scriptsize \fcolorbox{black}{amethyst}{\color{white}{TODO: #1}}}}}
\newcommand{\needscitation}[1][]{\textbf{\emph{\scriptsize \fcolorbox{black}{amethyst}{\color{white}{cite}}}}}

\fi

\newif\ifsubmission
\submissiontrue
\ifsubmission

\newcommand{\loiss}[1]{\textcolor{black}{#1}}

\newcommand{\todo}[1][]{}
\newcommand{\ntodo}[1][]{}
\newcommand{\figplace}[1][]{}
\newcommand{\loiscomment}[1]{}

\else

\newcommand{\ntodo}[1][]{\textbf{\emph{\scriptsize \fcolorbox{black}{amethyst}{\color{white}{TODO}}}}}

\definecolor{dollarbill}{rgb}{0.52, 0.73, 0.4}

\definecolor{ups-truck}{rgb}{0.53, 0.28, 0.21}

\definecolor{blue(ncs)}{rgb}{0.0, 0.53, 0.74}

\newcommand{\loiss}[1]{\textcolor{magenta}{#1}}
\newcommand{\loiscomment}[1]{\textcolor{red}{\textbf{[@lois: {\it#1}]}}}

\newcommand{\todo}[1][]{\textbf{\scriptsize \fcolorbox{black}{red}{\color{white}{TODO}}}}

\newcommand{\figplace}[1][]{\textbf{\scriptsize \fcolorbox{black}{blue}{\color{white}{FIGURE}}}
\textcolor{red}{\underline{$\overline{\hbox{\emph{#1}}}$}}}

\fi

\usepackage[font={small,bf}]{caption}

\usepackage{titlesec}

\makeatletter
\g@addto@macro{\normalsize}{%
   \setlength{\abovedisplayskip}{3pt plus 0.5pt minus 1pt}
  \setlength{\abovedisplayskip}{3pt plus 0.5pt minus 1pt}
  \setlength{\belowdisplayskip}{3pt plus 0.5pt minus 1pt}
  \setlength{\abovedisplayshortskip}{0pt}
  \setlength{\belowdisplayshortskip}{0pt}
  \setlength{\intextsep}{4pt plus 1pt minus 1pt}
  \setlength{\textfloatsep}{4pt plus 1pt minus 1pt}
  \setlength{\skip\footins}{5pt plus 1pt minus 1pt}}
\makeatother

\titlespacing\section{0pt}{2pt plus 1pt minus 1pt}{3pt plus 1pt minus 2pt}
\titlespacing\subsection{0pt}{2pt plus 1pt minus 1pt}{3pt plus 1pt minus 2pt}
\titlespacing\subsubsection{0pt}{2pt plus 1pt minus 1pt}{3pt plus 1pt minus 2pt}

\newcommand{\mechanism}{DR-STRaNGe}

\newcommand{\affilETH}[0]{\textsuperscript{\S}}
\newcommand{\affilETU}[0]{\textsuperscript{$\dagger$}}

\title{\mechanism{}: End-to-End System Design\\for DRAM-based True Random Number Generators} 
\author{{F. Nisa Bostancı\affilETU\affilETH}\qquad%
{Ataberk Olgun\affilETU\affilETH}\qquad
{Lois Orosa\affilETH}\qquad 
{A. Giray Ya\u{g}l{\i}k\c{c}{\i}\affilETH}\qquad \\%
{Jeremie S. Kim\affilETH}\qquad%
{Hasan Hassan\affilETH}\qquad%
{O\u{g}uz Ergin\affilETU}\qquad%
{Onur Mutlu\affilETH}\qquad\vspace{-3mm}\\\\
{\qquad \affilETU \emph{TOBB University of Economics and Technology} \qquad
\affilETH \emph{ETH Z{\"u}rich}} }

\begin{document}
\bstctlcite{IEEEexample:BSTcontrol}

\maketitle
\fancypagestyle{plain}{%
\fancyhf{} %
\fancyfoot[C]{\fontsize{12pt}{12pt}\selectfont\thepage} %
\renewcommand{\headrulewidth}{0pt}
\renewcommand{\footrulewidth}{0pt}}

\pagestyle{plain}
\thispagestyle{plain}

\setstretch{0.9}
\input{sections/1_abstract}

\input{sections/2_n_intro}
\input{sections/3_background}

\input{sections/4_n_motivation}
\input{sections/5_designchallenges}
\input{sections/6_1_mechanism}

\input{sections/6_3_interface}
\input{sections/6_4_security_analysis}
\input{sections/7_1_methodology}

\input{sections/8_1_perf_analysis}

\input{sections/8_2_fairness_analysis}
\input{sections/8_3_buffer_analysis}

\input{sections/8_4_scheduler_analysis}
\input{sections/8_5_pred_analysis}
\input{sections/8_6_quac_analysis}
\input{sections/8_7_other_analysis}

\setstretch{0.86}
\pagebreak
\input{sections/9_relatedwork}
\input{sections/10_conclusion}
\section*{Acknowledgments}
We thank the anonymous reviewers of MICRO 2021 and HPCA 2022 for feedback. {We thank} the SAFARI group members for feedback and the stimulating intellectual environment. We acknowledge the generous gifts provided by our industrial partners: Google, Huawei, Intel, Microsoft, VMware.
\setstretch{0.845}

\bibliographystyle{IEEEtranS}
\bibliography{ref}
\pagebreak
\onecolumn
\setstretch{1.0}
\appendix
\input{sections/11_appendix}

\end{document}

%% file: sections/1_abstract.tex
\begin{abstract}
Random number generation is an important task {in} a {wide variety of} {critical} applications including {cryptographic algorithms}, scientific simulations, {and} industrial testing tools. 
{True Random Number Generators (TRNGs) produce {cryptographically-secure} truly random {data} by sampling a physical entropy source that {typically requires custom hardware and suffers from long latency. T}o enable high-bandwidth  {and} low-latency TRNGs on widely-available commodity devices, recent works propose hardware TRNGs that generate random numbers using commodity DRAM as a{n entropy source}. {Although prior works demonstrate promising TRNG mechanisms using DRAM, practical integration of such mechanisms {into} real systems {poses various challenges.}}}

{We identify three key challenges for using DRAM-based TRNGs in current systems: (1)~generating random numbers with DRAM-based TRNGs can degrade overall system performance by slowing down concurrently-running applications due to the interference between RNG and regular memory operations in the memory controller (i.e., \emph{RNG interference}), (2)~this RNG interference can degrade system fairness by causing unfair prioritization of applications that intensively use random numbers (i.e., \emph{RNG applications}), and (3)~RNG applications can experience significant slowdown due to the high latency of DRAM-based TRNGs.}

To address these challenges, we propose \mechanism{}, an end-to-end system design for DRAM-based TRNGs that (1)~reduces {the RNG interference by separating RNG requests from regular memory requests in the memory controller}, (2)~{improves} fairness across applications with an RNG-aware {memory request} scheduler, and (3)~hides the {large TRNG latencies}
using a random number buffering mechanism combined with a {new} DRAM {idleness} predictor {that accurately identifies idle DRAM periods}.

We {evaluate} \mechanism{} {using} {a comprehensive set of 186 multi-programmed} workload{s}. {Compared to an RNG-oblivious baseline system, \mechanism{}} improves the performance of non-RNG and RNG applications on average {by} 17.9\% and 25.1\%, respectively. \mechanism{} improves {system} fairness by 32.1\% on average when {generating} random numbers {at} a $5\ Gb/s$ {throughput}. {\mechanism{} reduces energy consumption by $21\%$ compared to the RNG-oblivious baseline design by reducing the time spent for RNG and non-RNG memory accesses by $15.8\%$.} 
\end{abstract}

%% file: sections/2_n_intro.tex
\section{Introduction}
{R}andom numbers are {used} in {a wide range of applications}{,} such as cryptograph{ic} key generation, authentication, scientific simulations{,} Monte Carlo {methods}, {and} industrial testing ~\cite{bagini1999design,barangi2016straintronics,cherkaoui2013very,pareschi2006fast,gutterman2006analysis,von2007dual, kim2017nano,drutarovsky2007robust,kwok2006fpga,zhang2017high,quintessence2015white}. 
{These applications often require a high-throughput random number generator to achieve {high} performance{~\cite{drange,olgun2021quactrng,wang2016theory}}}.

{Random number generators are categorized into two classes ~\cite{tilborgencyclopedia,chevalier1974random, knuth1998art, tsoi2003compact}{.}
First, true random number generators (TRNGs)~\cite{pyo2009dram,keller2014dynamic,sutar2018d,hashemian2015robust,tehranipoor2016robust,wang2012flash,ray2018true,holcomb2007initial,holcomb2009power,van2012efficient,chan2011true,tzeng2008parallel,teh2015gpus,majzoobi2011fpga,wieczorek2014fpga,chu1999design,amaki2015oscillator,mathew20122,brederlow2006low,tokunaga2008true,bucci2003high,bhargava2015robust,kinniment2002design,holleman20083,gutterman2006analysis,dorrendorf2007cryptanalysis,lacharme2012linux,pareschi2006fast,yang2016all} harness the entropy resulting from inherently random physical processes (e.g., electrical noise, clock jitter, Brownian motion, and atmospheric noise) to generate random numbers. Second, pseudo-random number generators (PRNGs)~\cite{matsumoto1998mersenne,blum1986simple,mascagni2000algorithm,steele2014fast,marsaglia2003xorshift}} {use a seed value to produce a deterministic stream of numbers that {appears to be} random without the knowledge of the seed.}

{Random number quality} is critical for security applications such as authentication and key generation{~\cite{kocc2009cryptographic,gutterman2006analysis,von2007dual,kim2017nano,drutarovsky2007robust,kwok2006fpga,cherkaoui2013very,zhang2017high,quintessence2015white,clarke2011robust,lu2015fpga,hull1962random,ma2016quantum,botha2005gammaray,davis1956some}}. Since {PRNGs output a deterministic stream of numbers, {they are not preferable for} security{-}critical applications}~\cite{corrigan2013ensuring, von2007dual, cherkaoui2013very}.
{Instead,} {these applications} {use} TRNGs{,} because unlike PRNGs{,} TRNGs do {not} rely on {predictable} seed values that can
{compromise {application} security}.

{DRAM is widely available in almost all computer systems and can be {easily} integrated into mobile and IoT devices {as main memory}. To enable low-cost and high-throughput true random number generation in almost all commodity systems and emerging processing-in-memory (PIM) architectures, prior works~\cite{keller2014dynamic, sutar2018d,hashemian2015robust,tehranipoor2016robust,eckert2017drng,drange,bmstalukdertrng,olgun2021quactrng} propose various DRAM-based TRNGs that exploit manufacturing process variations in DRAM and the resulting entropy
to generate true random numbers. However, no prior work provides an end-to-end system design to enable DRAM-based true random number generation in real systems.}

{We identify \textbf{three key challenges} these proposals face {in} an end-to-end system design {that integrates a DRAM-based TRNG}.}
{First,} {performing} random number generation in DRAM {can be} {intrusive} in current systems that use DRAM as main memory. It can cause {a} significant {slowdown} on concurrently-running applications due to the interference {between RNG and non-RNG memory requests} {{in} the memory {{request} scheduler}} { (i.e., RNG interference)}.
{{Second}, existing memory schedulers {schedule memory requests to achieve high system fairness and performance}. However, they are RNG-oblivious and {ignore different} characteristics {of RNG and non-RNG memory requests. RNG requests are received in bursts and served together because {{interleaving RNG and non-RNG requests induce additional overhead from frequently modifying timing parameters}}. Existing memory schedulers can prioritize RNG requests to achieve high throughput {to serve} the high RNG demand. This can increase the memory stall time of non-RNG applications more than RNG applications and degrade system fairness.}}
{{Third}, random number generation in DRAM has a high latency and {can} cause applications that use random numbers intensively (i.e., \emph{RNG applications}) to experience long memory stall times. DRAM-based TRNG mechanisms need to perform multiple {DRAM} reads to gather {enough} random bits {from DRAM}. Random number generation can stall the {processor's} instruction window if later instructions depend on the generated random number. Therefore, RNG applications {can} suffer from {long} memory stall {times} even when they are the only application running on the system}.

\textbf{Our goal} is to design an end-to-end system for DRAM-based TRNGs with low cost and high performance that (1) minimizes the slowdown {of {both} RNG and non-RNG applications {by reducing and controlling} the interference between them}
, (2)~{improves system fairness by reducing the memory stall time experienced by non-RNG applications due to random number generation}, and (3)~{mitigates the performance degradation of RNG applications {due to} the} high latency of DRAM TRNG mechanisms. 
To this end, we propose \mechanism{}, a {new} end-to-end \textbf{S}ystem design for \textbf{DR}AM{-based} \textbf{T}rue \textbf{R{a}}ndom \textbf{N}umber \textbf{G{e}}nerators. 
\mechanism{} {has} three {major} components. 
{First, {\mechanism{} implements} a {\emph{buffering mechanism}} {to reduce} both the interference between RNG and non-RNG applications and
the high latency of {a} DRAM-based {T}RNG. 
Our buffering mechanism {comprises} (1)~a DRAM {idleness} predictor that leverages {a DRAM channel's idle time}
to generate random numbers with low interference {to the system}, and (2)~a random number buffer in the memory controller to store the generated values {in advance to mitigate {the} TRNG latency.}}
{Second, {\mechanism{} implements} a{n} {\emph{RNG-aware scheduler}} that improves system fairness and reduces memory stall time by reducing the {RNG} interference. {Our} RNG-aware scheduler uses a{n} RNG request queue {to} {separate {RNG requests} from {non-RNG requests}}, and prioritizes requests based on the priority of the applications set by the operating system {(OS)}. Third, {\mechanism{} provides} an \emph{{interface to applications}} that enables them to use the system's DRAM-based TRNG {with low latency and high throughput}.} 
\mechanism{} is independent of the DRAM-based TRNG mechanism used in the system, and is compatible with {all} previously proposed {DRAM-based TRNGs}.

{\textbf{Summary of Results.} Our experimental evaluations across a variety of real workloads show that }{\mechanism{} improves (1) { performance of both} non-RNG and RNG applications by 17.9\% and 25.1\% on average, respectively, compared to {the} RNG-oblivious baseline design and (2) system fairness by 32.{1}\% on average when an RNG application with a $5\: Gb/s$ random number {generation} throughput {requirement} runs {concurrently} with non-RNG applications.} 
\mechanism{} {provides} 21\% energy reduction {over} the RNG-oblivious baseline design by reducing the {total} time spent for {RNG and non-RNG} memory accesses by 15.8\%. {\mechanism{} incurs {a} minor area overhead {of} 0.0022$mm^2$ {at $22nm$ technology node} (0.00048\% of an Intel Cascade Lake CPU Core~\cite{wikichipcascade}).}
{This paper makes the following key contributions: 
\begin{itemize}
    \item \mechanism{} is the first work that {overcomes three {key} challenges {of DRAM-based TRNGs} and} proposes a viable end-to-end system design for DRAM-based TRNGs. 
    \item {We show that an efficient random number buffering mechanism can hide the high TRNG latency and reduce RNG interference on the system. We propose the first random number buffering mechanism with a lightweight DRAM {idleness} predictor that can predict idle periods with high accuracy {(8{0}\%)}. \mechanism{} generates random bits {during} predicted idle periods and fills the buffer, {such that, random numbers, { when requested,} {are served} with low latency {from the random number buffer}}.} 
    \item We propose {the} RNG-Aware {memory request} scheduler, a priority-based scheduler design that improves {system} fairness and reduces memory stall time by reducing the {RNG} interference. The scheduler achieves this by differentiating RNG {requests} from regular requests and employ{ing} different scheduling techniques {for them} to minimize the slowdown {of} high-priority applications.
    \item We evaluate {the} performance, fairness, and {energy consumption} of \mechanism{}, across a variety of real workload mixes, showing significant performance{,} fairness {and energy} benefits over the {baseline} RNG-oblivious {system} design. {\mechanism{} improves performance of non-RNG applications{,} system fairness{, and energy} by 17.9\%{,} 32.{1}\%{, and 21\%,} respectively{,} compared to the {commonly-used baseline} memory request scheduler design.}
    \item We evaluate \mechanism{} using two state-of-the-art DRAM-based TRNG mechanisms{:} D-RaNGe \cite{drange} and QUAC-TRNG \cite{olgun2021quactrng}. We show that \mechanism{} is compatible with these mechanisms and improves {system} performance{,} fairness {and energy with both TRNG mechanisms}. 
\end{itemize}}

%% file: sections/3_background.tex
\section{Background}
\label{background}

{{W}e provide a brief background on DRAM organization, memory schedulers, and DRAM-based TRNGs.}

\textbf{DRAM Organization.}
{DRAM-based main memories are 
accessed {via} a memory {channel} which is internally connected to multiple \emph{banks.} {E}ach {bank} contains a two dimensional array of DRAM cells, organized as \emph{rows} and \emph{columns}. {Upon a request,}
DRAM cells are fetched at row granularity and {the fetched row is} temporarily stored in a buffer (\emph{row buffer}). A request is serviced faster if the {requested} data is already in the row buffer (i.e., \emph{row buffer hit})~\cite{keeth2001dram}.} {For more information on DRAM, we refer the reader to {extensive} prior work~\cite{kim2012case, lee2013tiered,lee2015adaptive,seshadri2013rowclone,bhati2015flexible,chang_understanding2017,chang2016understanding,chang2014improving,chang2016low,chang2017understanding,vampire2018ghose,hassan2016chargecache,hassan2017softmc,khan2014efficacy,khan2016parbor,khan2017detecting,kim2018solar,kim2018dram,kim2014flipping,kim2016ramulator,lee2016reducing,lee-sigmetrics2017,lee2015decoupled,liu2013experimental,raidr,mukundan2013understanding,patel2017reaper,qureshi2015avatar,seshadri2016simple,seshadri2017ambit,zhang2014half}.}

\textbf{Memory {Request} Scheduling.}
{The} First Ready First Come First Serve (FR{-}FCFS)~\cite{rixner2000memory,zuravleff1997controller} scheduling policy prioritizes {1)} row hits over all other requests, {2)} then older requests {over} the younger ones. 
{FR-FCFS policy aims {to} maximiz{e} throughput}
by leveraging the row buffer.
However, it unfairly prioritizes the applications that have high {row locality and high {memory intensity}}~\cite{moscibroda2007memory, mutlu2007stall}.

{Previous work{s ~\cite{mutlu2009parallelism,kim2010atlas,kim2010thread,ghose2013criticality,ebrahimi2011parallel, subramanian2014blacklisting,mutlu2007stall,nesbit2006fair,rafique2007effective,usui2016dash,ausavarungnirun2012staged,subramanian2016bliss}} propose application-aware memory {request} schedul{ers, {which} improv{e} both} performance and fairness across applications.
{These schedulers} monitor different characteristics and schedule based on a maintained ranking {~\cite{mutlu2009parallelism,kim2010atlas,kim2010thread,ghose2013criticality,mutlu2007stall,ebrahimi2011parallel}} or use blacklisting ~\cite{subramanian2016bliss,subramanian2014blacklisting} based on {application} behavior.}
{Ranking based schedulers can unfairly slow down low-ranked applications, and have significant hardware complexity. {The}  blacklisting scheduler (BLISS) \cite{subramanian2016bliss,subramanian2014blacklisting} is a simple memory scheduler design that categorizes applications into only two groups based on sensitivity to inter-application interference. No prior work considers the high latency of random number generation or requests with different latencies due to non-standard timing parameters.}

\textbf{DRAM-based True Random Number Generators.}
Prior {research} proposes {various DRAM-based} TRNGs that leverage the randomness in DRAM timing failures~\cite{drange,bmstalukdertrng,olgun2021quactrng}, retention failures~\cite{keller2014dynamic,sutar2018d,hashemian2015robust} and start-up values~\cite{tehranipoor2016robust,eckert2017drng}. 

\pagebreak

Retention failure- and start-up value-based TRNGs are limited in the throughput they can provide~\cite{drange} {because r}andom retention failures can take minutes to accumulate {at} room temperature~{\cite{patel2017reaper,raidr,qureshi2015avatar,liu2013experimental,halderman2009lest,venkatesan2006retention}} and random start-up values are formed following {expensive} DRAM power-cycles~\cite{jedec2012}. {As such, t}hese mechanisms cannot be used in a streaming manner to generate true random numbers. 

DRAM TRNG mechanisms based on timing failures can be used in {a} streaming manner, and they provide {random numbers at} high throughput ($>100\ Mb/s$). {These TRNGs operate by deliberately violating DRAM timing parameters that are defined by the DRAM manufacturers to guarantee correct operation in DRAM arrays when obeyed by the memory controller.} Random DRAM timing failures can be induced quickly using carefully engineered {and timed} valid sequences of DRAM commands {on commodity DRAM devices~\cite{drange,bmstalukdertrng,olgun2021quactrng}}. 

{To generate random numbers in DRAM, prior work \cite{drange} reserves rows that contain RNG cells, which can be used to extract random numbers by reading with reduced timing parameters. The throughput of the TRNG depends on the number of RNG cells per row. The latency of the TRNG depends on the latency of the DRAM command sequence used to induce random errors. Random number generation can be an {intrusive} procedure {to system operation, especially} when the required random number {throughput} is high, {due to} {two reasons{.}} {First, random number generation} requires multiple reads to multiple reserved rows. {{S}econd, due to non-standard timing parameters, DRAM becomes unavailable for regular memory requests} to ensure {reliability} of data residing in other rows. }

%% file: sections/4_n_motivation.tex
\section{Motivation and Goal} 
\label{Motiv}

TRNGs are used in a wide range of security applications such as cryptographic key generation, authentication, {generation of} {initial and} data padding values, and countermeasures against hardware attacks { ~\cite{gutterman2006analysis,von2007dual,kim2017nano,drutarovsky2007robust,kwok2006fpga,zhang2017high,quintessence2015white,cherkaoui2013very,kim2014flipping}}. {The} quality of random numbers is important to ensure system security in presence of attacks that aim to obtain confidential user data~\cite{von2007dual,cherkaoui2013very}. Emerging security protocols (e.g., quantum key distribution protocols~\cite{clarke2011robust,lu2015fpga}) provide stronger security guarantees in {the} presence of attacks {targeting weak random numbers}. These protocols require {very high {true} random number} throughput, {in} the order of several $Gb/s$ \cite{wang2016theory}. 

High throughput DRAM-based TRNGs have the main advantage of availability over {other TRNGs that typically require} dedicated hardware. DRAM devices are widely available in most computer systems and {in emerging processing-in-memory architectures ~\cite{mutlu2020modern,ghose2019processing,mutlu2019processing,singh2021fpga,ghose2018enabling,wang2020figaro,giannoula2021syncron,akin2015data, aga2017compute, ahn2015scalable, ahn2015pim, lee2015simultaneous, seshadri2015fast, seshadri2013rowclone, seshadri2015gather, seshadri2017ambit, liu2017concurrent, seshadri2017simple, pattnaik2016scheduling, babarinsa2015jafar, farmahini2015nda, gao2015practical, gao2016hrl, hassan2015near, hsieh2016transparent, morad2015gp, sura2015data, zhang2014top, hsieh2016accelerating, boroumand2017lazypim, chang2016low, kim2018grim, ghose2018enabling, boroumand2021mitigating, boroumand2018google, seshadri2016simple, mutlu2019processing, CROW, singh2020near, fernandez2020natsa, kwon202125, devaux2019true, li2016pinatubo, chi2016prime, orosa2019dataplant, orosa2021codic}. }
{DRAM-based TRNGs can provide true random numbers to security-critical applications that run on these systems at high throughput. The impact of integrating DRAM-based TRNGs to these systems is twofold. First, DRAM-based TRNGs can enable security applications on mobile and IoT devices{,} which becomes more critical with increasing demand for user data privacy. Second, for PIM architectures, DRAM-based TRNGs can improve (1) system efficiency by enabling large contiguous code segments to be executed in memory, and (2) system security by enabling security tasks to run {completely} in memory.}

{Previous research proposes a set of high-throughput DRAM-based TRNGs. However, no prior work provides an end-to-end system design}
for integrating DRAM-based TRNGs in{to} real systems. We identify three {key} challenges for integrating DRAM-based TRNGs into a {baseline} \emph{RNG-oblivious} real system:
{(1) the interference of RNG and non-RNG applications {in} the memory controller 
causes {unnecessary} slowdowns {for} both {types of applications}, {(2)}  the unfair prioritization of RNG applications {that require high TRNG throughput} degrades {system fairness}, and {(3)} the high latency of DRAM-based TRNGs {degrades} the RNG {applications'} performance.}

To demonstrate the impact of using DRAM-based TRNGs in a{n} RNG-oblivious system, we simulate a realistic two-core system with a DRAM-based TRNG mechanism. The RNG-oblivious system generates 64-bit random numbers by changing DRAM timing parameters to induce errors in the reserved rows{\cite{drange}} as applications request random values. During random number generation, the system stalls regular memory requests because changed timing parameters can induce errors on real data residing in other rows. The system uses all memory channels in parallel to achieve the minimum {RNG} latency, and minimizes the time that regular requests are stalled. It is possible to use {only} one channel at a time. {However, {since}} the system does not know which channels will be idle at any given time, this can cause further slowdowns by blocking one busy channel for a longer time period. Using all channels and all banks is important to minimize the 
interference {of RNG} while serving the random number request{s} as quickly as possible.

{For our tests,} we create {172} two-core workloads{,} each consisting of one RNG and one non-RNG application. 
For RNG applications, we use {four different} synthetic benchmarks that request random numbers with {required RNG throughput values of} $640 Mb/s$, {$1280Mb/s$, $2560Mb/s$, and} $5120 Mb/s$. 
For non-RNG applications, we use 43 single-core applications from the following benchmark suites: SPEC CPU2006 \cite{spec2006}, TPC \cite{tpcweb}, STREAM \cite{McCalpin2007}, MediaBench \cite{fritts2009media}, and YCSB benchmark {suite}~\cite{ycsb}. 

\begin{figure}[h]
\includegraphics[width=\linewidth]{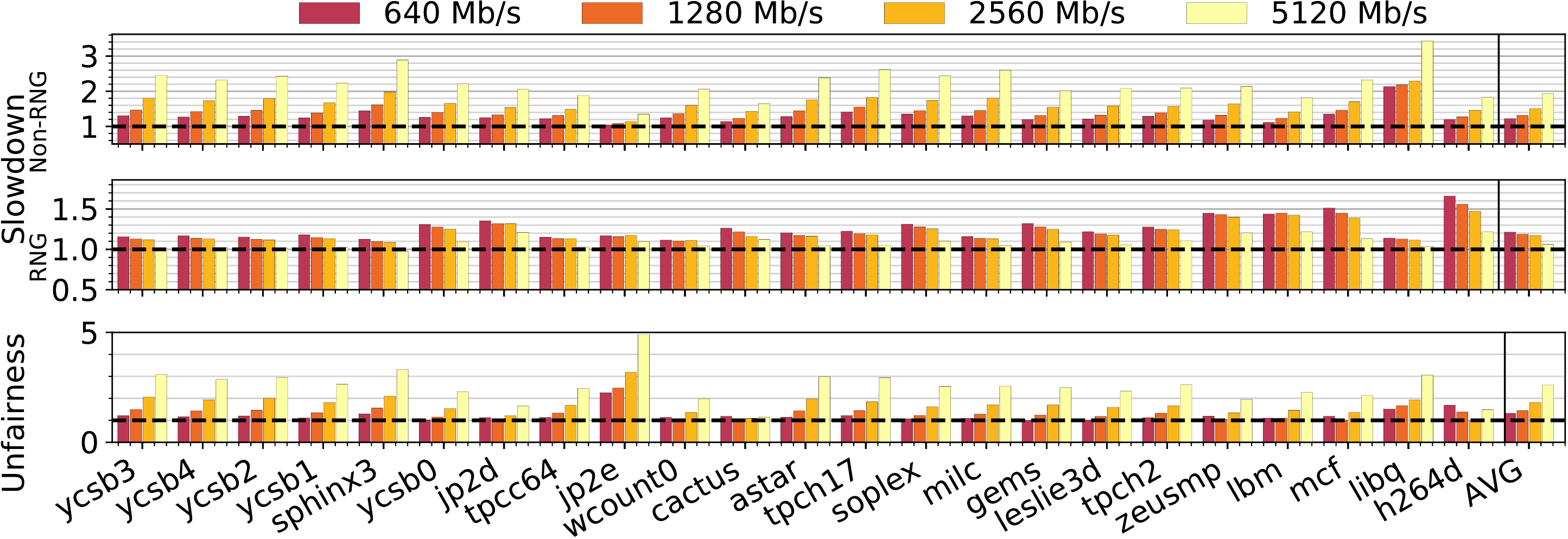}
\caption{{Slowdown} of non-RNG (top) and RNG (middle) applications running concurrently and the unfairness index of the two-core system (bottom) for {various} {RNG throughput requirements}. {AVG is across 172 workloads.}}
\label{fig:motiv3}
\end{figure}

{Figure \ref{fig:motiv3} (top and middle) shows the execution time of non-RNG and RNG applications running concurrently, normalized to {each} application running alone. {Figure~\ref{fig:motiv3}} (bottom) shows the unfairness index of the overall system for the same two-core workloads. }Unfairness index is calculated as the ratio of the maximum memory-related slowdown experienced by an application in the workload to the minimum memory-related slowdown~\cite{mutlu2007stall,gabor2006fairness,moscibroda2007memory}, explained in detail in Section \ref{methodology}. {An u}nfairness index of 1 means all applications experience {the same} memory-related slowdown. Higher unfairness index values indicate that one or more applications are unfairly prioritized by the {memory} scheduler.

\pagebreak
{\textbf{Impact of the Interference.} Figure \ref{fig:motiv3} (top and middle) shows that both {non-RNG and RNG applications {experience significant} slowdown{s due to} the interference of both types of applications in the memory scheduler.}

{The slowdown of non-RNG applications increases {as} the {required RNG throughput increases.}}
On average, non-RNG applications {experience} 93.1\% slowdown when {the required RNG throughput is {$5\:Gb/s$}}.} 
Figure \ref{fig:motiv3} (middle) shows that RNG applications experience slowdowns due to sharing the main memory with another application. {W}e show that the least and the most RNG intensive applications {experience} 21.4\% and 6.2\% {average} slowdown{, respectively,} {compared to when they run alone}.

{\textbf{Unfair Prioritization.} Figure \ref{fig:motiv3} (bottom) shows the unfairness index. We show that the unfairness indices of applications grow significantly with increasing RNG throughput requirement. On average, workloads have an unfairness index of $1.32$ at an RNG throughput requirement of $640Mb/s$ and it gradually increases to $2.61$ when the required RNG throughput increases to $5120 Mb/s$.} 

{\textbf{Impact of the RNG Latency.} {W}e observe that the RNG applications spend up to $58.8\%$ of their execution time in {random number generation} when {they require high RNG throughput}, {due to} the high RNG latency.}

\begin{figure}[h]
\centering
\includegraphics[width=0.95\linewidth]{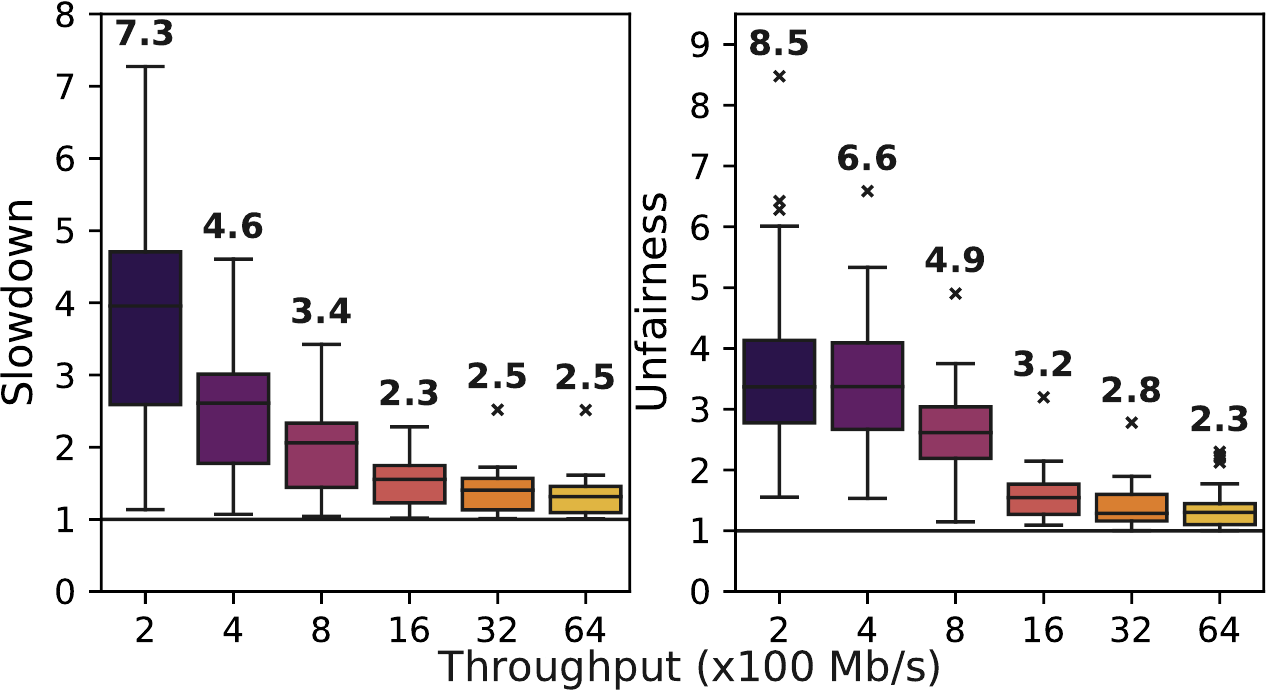}
\caption{The effect of DRAM TRNG throughput on non-RNG application performance (left) and system fairness (right).}
\label{fig:throughput-motivation}
\end{figure}

{{\textbf{Impact of the TRNG Throughput.} W}e simulate 6 different DRAM-based TRNG mechanism{s} with {different} TRNG throughput values {ranging from $200\:Mb/s$ to $6.4\:Gb/s$. Note that} two state-of-the-art DRAM-based TRNGs, D-RaNGe~\cite{drange} and QUAC-TRNG~\cite{olgun2021quactrng}, {can provide  ${\sim}563Mb/s$ and ${\sim}3.44Gb/s$ average {random number} throughput{, respectively,} given a state-of-the-art system configuration~\cite{olgun2021quactrng}}.\footnote{
All designs assume low latency values
based on D-RaNGe’s latency{~\cite{drange}} to show only the effect of TRNG throughput. Note that QUAC-TRNG’s latency is higher than
what is assumed in this figure. Therefore, the real slowdown
and unfairness indices of a system with QUAC-TRNG{~\cite{olgun2021quactrng}} would
be higher.}} {We evaluate 43 two-core multi-programmed workloads, each consisting of one RNG and one non-RNG application. Figure~\ref{fig:throughput-motivation} plots the distribution of (1) the slowdown of non-RNG applications {(left)}, and (2) the system fairness {(right) across all 43 workloads}.}

Each box in the figure represents the interquartile range of the observed slowdown ratios and unfairness indices of a dual-core baseline system {when the system is} provided with {the TRNG throughput displayed on the x-axis}. The midlines of the boxes indicate the median value of the corresponding interquartile ranges. {Cross ($\times$) marks represent the outlier values that fall outside the interquartile range (by more than 1.5$\times$ the length of the box away from the upper quartile).}
{We label the maximum observed slowdown and the unfairness index above each box in the figure.}

{We make two observations. First, workloads experience significant performance degradation and high system unfairness when executed on systems with state-of-the-art DRAM-based TRNGs. {Even the highest-throughput existing TRNG causes {a} {$39.9\%$} {average} slowdown and degrades {average} fairness by {$28.5\%$}.} Second, performance and system fairness do not improve significantly even with future DRAM-based TRNGs that can {potentially} provide higher throughput. {The maximum slowdown and unfairness we observe across all workloads {begin to} saturate {as TRNG throughput reaches $3.2\:Gb/s$}.}}

{\textbf{Our goal}} in this paper is to develop a {low-cost and high-performance} end-to-end {system design for DRAM-based TRNGs} that (1) reduces the interference between RNG and non-RNG applications, (2) improves system fairness across RNG and non-RNG applications, and (3) hides the high latency of DRAM TRNGs that RNG applications suffer from. 

%% file: sections/5_designchallenges.tex
\section{Design Opportunities}
In this section, we discuss the design opportunities of an end-to-end system for DRAM-based TRNGs.

\textbf{Random Number Buffering.}
Previous work \cite{drange, olgun2021quactrng} assumes the memory controller can generate random numbers when DRAM utilization is low and uses extra bandwidth {to fill a buffer of random numbers}. This assumption is not complete without a DRAM {idleness} predictor since the controller does not know for how many cycles {memory} channels will be available to generate random numbers with low or no overhead. A buffering mechanism combined with a DRAM {idleness} predictor is needed to {accurately} identify the idle periods where the system can generate random numbers {with minimal performance impact}.

\textbf{Memory {Request} Scheduling.}
{The latency of RNG is an important factor in system slowdown because it increases the memory stall time of non-RNG requests of both RNG and non-RNG applications. An RNG-aware scheduler is needed to control both {types of} applications' memory stall time{s} by managing when to generate random numbers.}

\textbf{Application Interface.}
The system needs to expose an interface {so that applications can communicate with the DRAM-based TRNG}. The interface can be designed in many ways, including memory-mapped configuration status registers {(CSRs){~\cite{wolrich2004mapping}}}{,} other existing I/O {interfaces} {(e.g., x86 IN and OUT {instructions})} {or} new specialized ISA instructions. {The interface} need{s} to be exposed to user application{s} {via} system calls or {static/dynamic library API functions}. 

%% file: sections/6_1_mechanism.tex
\section{\mechanism{}}

{\mechanism{} is a \om{high performance and low cost} end-to-end {DRAM-based TRNG} system{, which} \om{(1) reduces the RNG interference that degrades overall system performance, (2) improves the system fairness across RNG and non-RNG applications, and (3) hides TRNG latency \loiss{to improve the performance of RNG applications}}.}
\loiss{\mechanism{}} consists of three components. \loiss{First}, {its} \om{\emph{Random Number Buffering Mechanism}} (Section \ref{Random Number Buffer})
aims to hide the high {TRNG} latency \loiss{by} utilizing idle {DRAM cycles} to generate random numbers with low interference {on other running applications}. The buffering mechanism achieves this by predicting {and utilizing} the idle periods in {DRAM} channels {where it can} generate random numbers in small batches. The buffering mechanism then stores the generated {random numbers} in a small buffer in the memory controller {and serves incoming random number requests from this buffer}. \loiss{Second}, {the} RNG-aware Scheduler (Section \ref{RNGA Scheduler}) aims to reduce the memory stall time {{of }regular memory requests} by scheduling RNG requests {in an order that reduces the RNG interference and improves system performance.}
{The scheduler accumulates RNG and non-RNG memory requests in {separate} {memory request} queues and {chooses {from} which {queue} to schedule a request based on the priority levels of {concurrently-running} applications.}}
\loiss{Third}, {applications use} {\mechanism{}'s} Application Interface (Section \ref{App Interface}) {to utilize DRAM-based TRNGs}. \mechanism{} achieves its design goals by combining these three components. 

\begin{figure}[ht]
\centering
\includegraphics[width=0.85\linewidth]{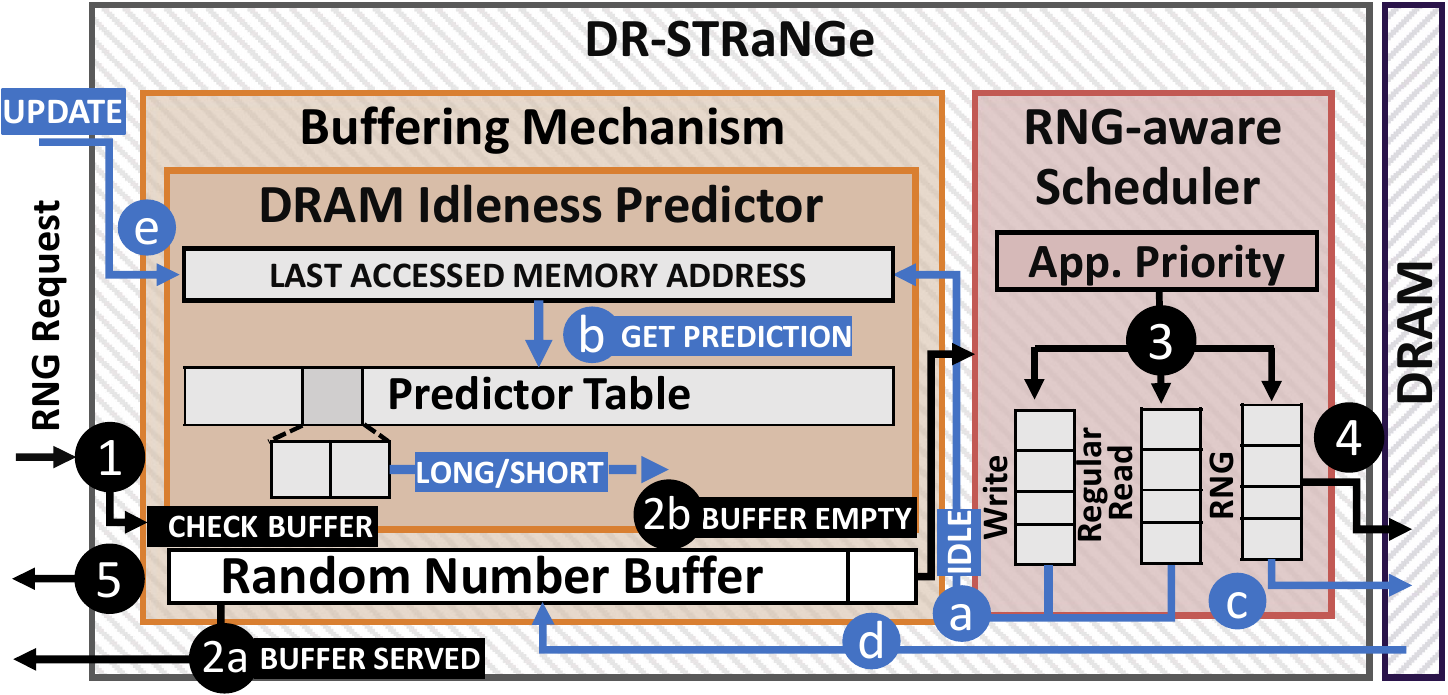}
\caption{\mechanism{} Overview.}
\label{fig:overview}
\vspace{-2mm}
\end{figure}

\begin{figure}[ht]
\centering
\includegraphics[width=0.8\linewidth]{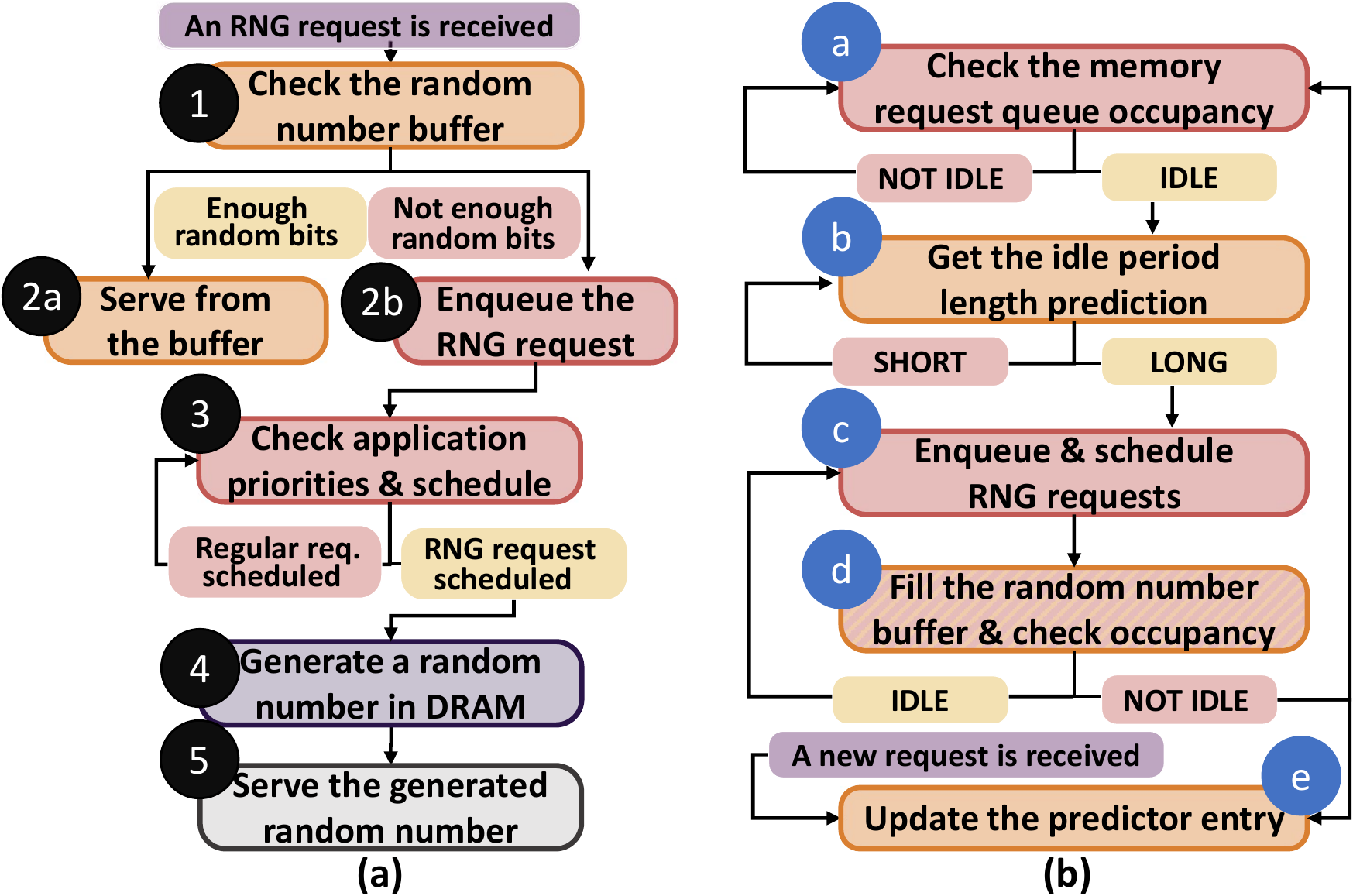}
\caption{\mechanism{} Flowchart.}
\label{fig:flowchart}
\vspace{-1.5mm}
\end{figure}

\loiss{\textbf{\mechanism{} overview.}} Figure \ref{fig:overview} shows {an} overview of \mechanism{}, where \loiss{(1) the black track shows the flow of a random number request, and (2) the blue track shows how \mechanism{} fills the {random number} buffer. {Figure~\ref{fig:flowchart} shows the flowcharts of these two tracks.}}

The mechanism works as follows. There are two different execution modes in the memory controller: 1) {\emph{Regular Execution Mode} processes} {only} regular memory requests, and 2) {\emph{RNG Mode} handles} RNG read requests and gathers random {numbers}. The {memory controller} is \loiss{initially} in {Regular} Execution Mode.

\loiss{When the memory controller receives} an RNG request, \mechanism{} checks the random number buffer {(\circled{1})}.
If the random number buffer has enough random bits, \mechanism{} serves the request from the buffer with low latency {(\circled{2a})}. {Otherwise,} {if} the buffer {does not have a sufficient number of random bits,} \mechanism{} {enqueues an RNG request to the memory request queue} (\circled{2b}). 
{{T}he} RNG-aware scheduler {determines the memory request scheduling order by checking the concurrently running applications' priorities} {(\circled{3})}. 
{{Before \mechanism{} schedules RNG requests, it {first} schedules the older regular read requests {that belong to} high-priority non-RNG applications}. {After these requests are scheduled}, the memory controller switches to the RNG mode, {schedules} RNG requests and generates random bits in DRAM (\circled{4}).}
When {enough random bits} are gathered, \mechanism{} serves the random number request {(\circled{5})} and {the memory controller} switches back to {the} {Regular} Execution Mode.

{Every} cycle, \mechanism{} {determines} the idleness of each {DRAM} channel {by checking the number of requests in {memory request} queues} (\textcolor{blue}{\circled{a}}). If a channel{'s {memory request} queues are empty}, \mechanism{} {predicts} the idle period length (\textcolor{blue}{\circled{b}}). {When the DRAM {idleness} predictor} {finds a sufficiently long idle period} to generate random bits with no or low overhead, \mechanism{} enqueues and schedules RNG requests (\textcolor{blue}{\circled{c}}). \mechanism{} switches to {the} RNG Mode and generates random bits {by} utilizing all banks of the selected channel (\textcolor{blue}{\circled{d}}). If the channel {remains} idle {after random number generation}, {\mechanism{}} continues {to} fill the {random number} buffer. Random number generation stops when there is a new regular read request to the selected channel {or when the buffer is full}{.}  \mechanism{} switches to the {Regular} Execution Mode and updates the predictor table  {corresponding to the selected channel}
(\textcolor{blue}{\circled{e}}). {{When} the random number buffer is full, \mechanism{} stops {generating random numbers for the random number buffer} until a random number is served from the buffer.}

\subsection{Random Number Buffering Mechanism}
\label{Random Number Buffer}

{The goal of the random number buffering mechanism is to hide the high RNG latency by generating random numbers before they are requested. This improves system performance by (1) reducing the memory stall time {experienced by} RNG applications due to RNG latency and (2) mitigating the {RNG} interference {in the memory controller}. The key idea is to {use idle DRAM cycles to} generate random numbers and store them in a small buffer in the memory controller. {When the random number buffer is not empty}, \mechanism{} {quickly} serves random number requests from the buffer with low overhead. {Many} applications often do not fully utilize the DRAM bandwidth~{\cite{lee2000eager,david2011memory,lee2010dram,stuecheli2010virtual}}, and {thus} \mechanism{} can use the idle {DRAM bus} cycles for RNG.} {We refer to multiple consecutive idle cycles as \emph{idle periods}. {L}engths of idle periods {differ between} application{s} based on the {observed} memory access pattern. }

{The main challenge of the buffering mechanism is {determining} the length of an {upcoming} idle period {because the future memory accesses of an application are unknown}.} {{However, several} heuristics can be employed to predict the idle period length. \mechanism{} uses two different approaches: (1) assuming every idle period is sufficient{ly long} to generate 8 random bits, and (2) {predicting the length of the next idle period} based on idle period length history.}

\begin{figure}[ht]
\includegraphics[width=0.95\linewidth]{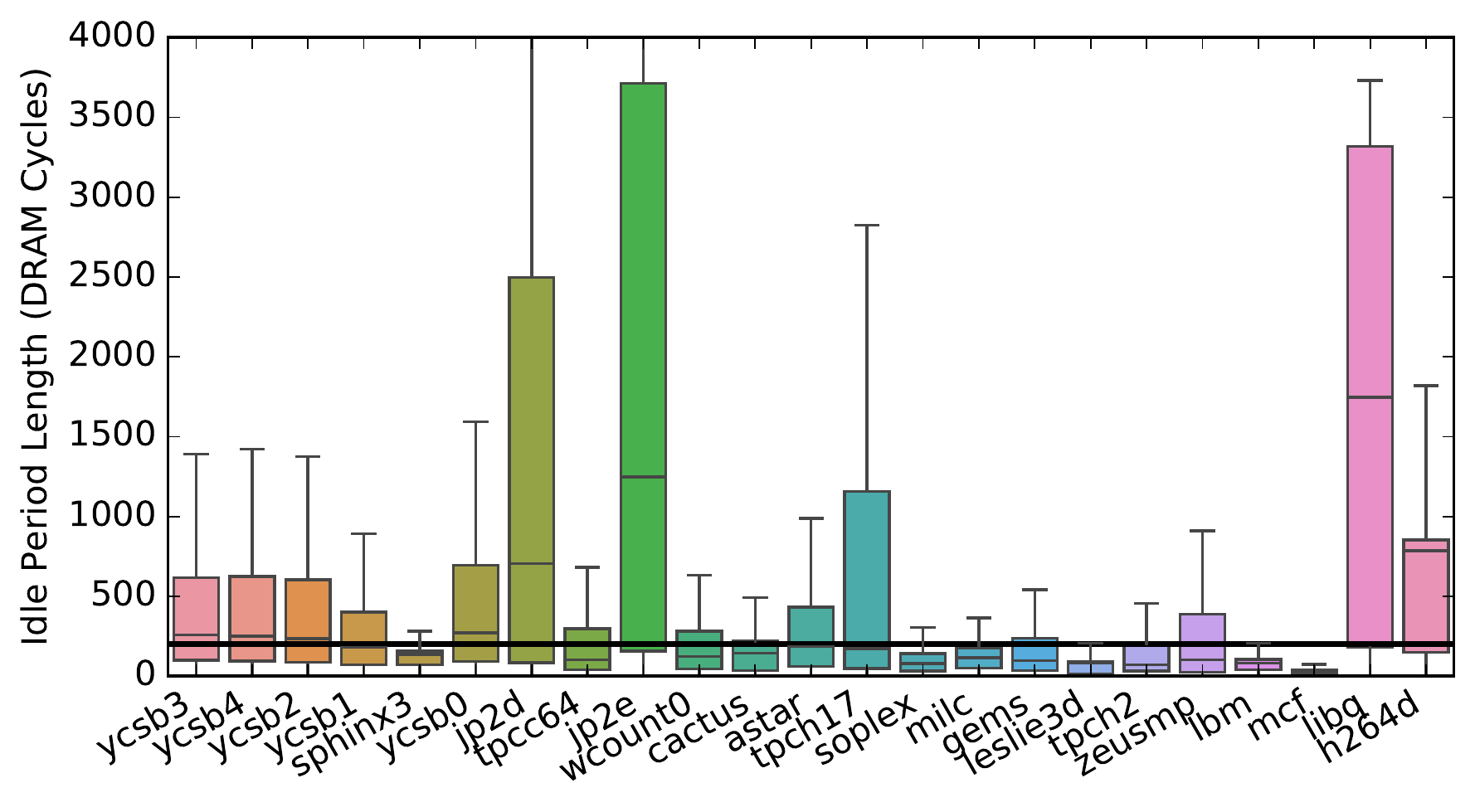}
\caption{{Distribution of} DRAM Idle Period Lengths}
\label{fig:idletime}
\end{figure}

{{Figure \ref{fig:idletime} plots the distribution of the idle DRAM period lengths of medium and high memory intensity applications {that we evaluate}}. {The distribution is represented as a box-and-whiskers plot where the y-axis {shows} the length of the observed idle DRAM periods in DRAM cycles. The {straight} horizontal line represents the time needed to generate a 64-bit random number, which is} {198 memory cycles ($990\:ns$) on average} in our test setup
{that {we describe} in {Section} \ref{methodology}}. }

{Each box represents the interquartile range of the observed DRAM idle periods and the middle line shows the median. We observe that, for many applications, a significant part of the idle periods do not pass the threshold of 198 cycles. {This means} that {the idle periods} are not sufficiently long to generate 64-bit random numbers. Therefore, {generating random numbers in smaller batches is preferable to achieve low interference with applications' memory accesses}.}
\pagebreak

{{We} design \mechanism{} to quickly generate {at least} eight random bits whenever a channel is idle. To leverage bank-level parallelism, we simultaneously access all DRAM banks and read {at least} one random bit from each bank during an idle period.}

{Generating random numbers in small batches does not solve the problem entirely {due to existence of} idle periods that are {even} shorter than the {time interval required to generate an 8-bit random number} {(40 cycles)}. {Generating random numbers during these short idle periods can degrade system performance by stalling regular {memory} requests.} Therefore, the mechanism needs to avoid {generating random numbers during} short idle periods. 
Our buffering mechanism consists of two parts: (1) a lightweight DRAM {idleness} predictor that identifies idle periods that {are at least 40 cycles long (i.e., \emph{{long idle periods}})} and (2) a small buffer that stores random numbers.} 
{Since the DRAM {idleness} predictor adds area and time complexity to the design, we evaluate {the random number} buffering mechanism with {and without} the {idleness} predictor in Section \ref{analysis}.}

\subsubsection{Simple Buffering Mechanism}
Our simple buffering mechanism {does \emph{not} {predict} for the length of {idle} DRAM period{s} and} leverages {\emph{every}} idle {cycle} {of a DRAM channel to generate random numbers {and fill the buffer}}. 

{With the} simple buffering mechanism{, \mechanism{}} selects {a} channel {for RNG} when {the} channel has no regular request in read and write queues. When a channel is selected, {\mechanism{} puts the channel into} {the} RNG Mode and gathers {at least} 8 random bits. {After generating at least 8 bits, {\mechanism{}}} {switches} back to {the Regular} Execution Mode {if the random number buffer is full or the memory scheduler has new regular memory requests}. {Regular} requests {that are at the memory scheduler} while the channel is in RNG mode are served after {switching} {to} the {Regular Execution Mode}.

\subsubsection{DRAM {Idleness} Predictor}
\label{predictor}
{The main goal of the DRAM {idleness} predictor is to accurately identify sufficiently long idle DRAM periods for RNG. {We design {two} DRAM idle period length predictors that differ in {their} prediction {mechanism{s},} accuracy and complexity.} }

{\textbf{Simple DRAM Idleness Predictor.} We propose a lightweight DRAM channel {idleness} predictor that uses last accessed memory addresses to predict idle period lengths.} {Our predictor {maintains} a table of 2-bit saturating counters, a register for \emph{last accessed address value}, {and} a counter for \emph{idle period length} initialized as 0 for each channel. A {channel}'s {predictor} table is accessed with the \emph{last accessed memory address} when its {memory request} queues are empty. We group idle periods into two categories: (1) \emph{{long}} (\# of cycles $\geq Period Threshold$) {and} (2) \emph{short} (\# of cycles $< Period Threshold$). 
{The predictor {treats} the idle {period} as \emph{{long}} if the last accessed address's counter is 2 or larger {and otherwise as \emph{short}}.}}

{During idle periods, \mechanism{} increments the \emph{idle period length} counter by one every cycle. When a channel receives a regular memory request, \mechanism{} updates the predictor table of the corresponding channel as follows. First, {\mechanism{}} retrieves the predictor table entry for the last accessed memory address to access the saturating counter. Second, if the observed idle period length is at least as long as the \emph{Period Threshold} (empirically determined as {40} cycles), {\mechanism{}} increments the saturating counter by one. Otherwise, it decrements the saturating counter. Finally, \mechanism{} sets the \emph{idle period length} to zero and updates the \emph{last accessed memory address} with the new request's address.}

The accuracy of the predictor affects \mechanism{}'s performance in two ways. First, if the predictor mispredicts {a} short idle period as {a} {long} {idle period}, which we refer to as {a false positive}, the interference between RNG and non-RNG applications {can increase}. Second, if the predictor mispredicts {a} {long} {idle} period as {a} short {idle period}, which we refer to as {a false negative}, it wastes an idle period; hence, {future random number requests {can experience the full} {RNG} latency {if the random number buffer is empty}.} {The predictor cannot starve RNG applications by predicting all idle periods as \emph{short} {because} \mechanism{} serves random number requests by generating random numbers on demand when the random number buffer is empty.}

We observe that the predictor often predicts idle periods as short due to the {large} number of short idle periods. This results in a {small} number of idle periods used for random number generation{. H}ence{,} \mechanism{} {provides} limited performance gain by using the simple DRAM {idleness} predictor. {To increase the RNG opportunities, we propose a method to utilize}
the periods when {a DRAM channel has low utilization} {(i.e., the memory request queue is largely empty)} {to generate random numbers for the buffer}.

{We {augment the predictor to determine {if a DRAM channel has low utilization} based on} a threshold, called {the} \emph{low utilization threshold}. {To determine if a DRAM channel has low utilization,} the predictor first}
checks the memory request queue occupancy and determines DRAM utilization {to be} low {if the} {the number of requests in the memory request queue is less than the {low utilization} threshold value {(empirically determined as 4)}.} Second, {when the DRAM utilization is low}, {the predictor} accesses {the last accessed memory address's entry in the predictor table} and predict{s} the length of the low utilization period. If the simple DRAM idleness predictor predicts the period {to be} {long}, \mechanism{} stalls the regular read queue and generates random numbers for the random number buffer. {The predictor} stalls only a {small number of} requests {as {the predictor does} not trigger RNG if there are more {requests} than the {\emph{low utilization threshold} value} in the request queue}. Some regular memory requests experience higher latency due to RNG but serving {the RNG} requests with lower TRNG latency improves the overall system performance.

{\textbf{Reinforcement Learning Agent for DRAM {Idleness} Prediction.}}
{{As a} second {predictor}, we design a reinforcement learning agent to predict the DRAM idle period lengths and create a model by {defining the DRAM idleness prediction} problem as a reinforcement learning problem.} 

Reinforcement learning (RL) is a commonly employed technique to solve architectural optimization problems such as branch prediction~\cite{zouzias2021branch}, memory scheduling~\cite{ipek2008self,mukundan2012morse}, prefetching{~\cite{peled2015semantic,bera2020pythia}}, dynamic voltage swing control for I/O communication~\cite{sai2016qlearning}, and garbage collection~\cite{kang2018dynamic}. Among RL techniques, Q-learning~\cite{Watkins92q-learning} is a preferable method due to its simplicity and model-free nature, making it a practical and effective solution. 
{A} Q-learning-based RL agent maintains a state machine, where performing a certain action ($a$) at a particular state ($s$) has a value, called Q-value, denoted as $Q(s,a)$. These Q-values represent the future cumulative reward of each action for each state and they are stored in a look-up table for  fast access.
At a given state ($s$), the algorithm chooses the action ($a$) with the largest Q-value, $Q(s,a)$.
We define two possible actions in our mechanism: 1) to initiate random number generation procedure and 2) to wait. In our implementation, the state is defined as the last accessed address whose least significant 10 bits are {{XOR}'ed} with the history of last 10 idle periods, where a logic-1 represents a \emph{{long}} idle period, {and} {a} logic-0 means a \emph{short} idle period.

After the agent takes an action (\emph{a}) based on the state (\emph{s}) of the environment, {the} Q-value of that state-action pair ($Q(s,a)$) is updated when the reward \emph{r} is {determined}. {Reward \emph{r} depends on the idle period length and the action taken according to the RL-agent's prediction and it is added to the Q-value of the state-action pair.} Positive rewards are used when the agent {correctly} predicts the {idle period length} and generates random numbers in {long} idle periods, or decides to wait in short idle periods. {In contrast, n}egative rewards are used when {the agent} mispredicts and causes more interference by initiating {the} RNG procedure in short idle periods {(false positive{s})} or waits in {long} idle periods {(false negative{s})}. {An action's reward} is calculated at the end of {the} idle period {when the agent can observe the idle period's length and determine the correctness of its prediction}. 

{T}he agent can observe the state {only} when {a DRAM channel} is idle. {The next state cannot be determined before taking an action because it depends on future memory accesses.} {Therefore, we omit the next state's part in the update function. T}he Q-value update function {is}  $Q(s,a) = (1 - \alpha)Q(s,a) + \alpha * r$ where $\alpha$ is the learning rate. {A high learning rate enables fast adaptation to changes in {the} access pattern but it is susceptible to noise. Based on our experiments, we observe that the agent achieves peak performance when the learning rate is $0.05$.}

\subsection{RNG-Aware {Memory Request} Scheduler}
\label{RNGA Scheduler}
{\mechanism{} cannot serve all random number requests from the buffer {due to the distribution of idle DRAM periods}. {Memory request s}cheduling becomes more critical when a random number is requested, the buffer is empty, and there are {RNG} and regular read requests waiting in the queues.} {The goal of the RNG-aware scheduler is to {(1)} schedule RNG requests without stalling other requests significantly and {(2)} improve system fairness.}

{RNG requests can {(1)} block regular memory requests when {a single memory request} queue is used for both types of requests and {(2) cause {the} memory controller to switch between the {Regular} Execution Mode and the RNG Mode frequently.} Since RNG and regular memory requests are different types of requests, it is intuitive to have separate queues for each {type}. \mechanism{} uses and maintains an additional queue, \emph{the RNG queue}, for RNG requests. {Two queues {reduce contention} for queue space between {the} two types of requests.}}

\subsubsection{Configuring The Scheduler Using Application Priorit{ies}} The operating system (OS) manages hardware resources based on priority levels of applications, and it is possible to use these priority levels for memory request scheduling. The RNG-aware scheduler can use these priority levels to prioritize either the RNG queue or the regular read queue. {However, these priority levels are insufficient to employ application-level scheduling decisions because the regular read queue is shared across all applications.
The scheduler needs to identify the applications that use both the RNG and regular read queues (i.e., \emph{RNG applications}). \mechanism{} marks an application as an RNG application when it requests a random number for the first time. After identifying RNG and non-RNG applications, the scheduler can use the priority bits set by the OS and prioritize one queue over the other.  The deprioritized queue can suffer from longer memory stalls, leading to starvation. The RNG-aware scheduler uses} a set of rules to {enforce priorities while} prevent{ing} starvation, explained as follows.

\textbf{RNG Prioritized.} {When an RNG application is prioritized over other applications, \mechanism{} prioritizes {the RNG application's} RNG requests to {(1) alleviate the slowdowns incurred due to RNG and (2) reduce the RNG interference.}} The scheduler chooses the RNG queue over the regular read queue when both queues are non-empty and {an} RNG application {with a request has higher priority than any} non-RNG application {with a request}. {The scheduler {chooses} the RNG queue until the last request in the queue is scheduled.}
When \mechanism{} completes the {{high-priority} random number generation {request}}, the scheduler starts to schedule requests from the regular read queue, preventing the starvation of non-RNG applications {with a request}. However, as a result of the {required }RNG {throughput}, the scheduler can choose the RNG read queue once it receives another {high-priority} RNG request.

\textbf{Non-RNG Prioritized.} The scheduler chooses the regular read queue over the RNG queue {and aims to minimize the memory stall time of non-RNG requests} when both queues are non-empty and {a} non-RNG application {with a request has} higher priority than {any} RNG application {with a request.} 
{The scheduler {only chooses} {the RNG} queue {when} the oldest request in the {regular} read queue is (1) from an RNG application, and (2) received after the oldest RNG request in the RNG queue. {In this case, t}he scheduler {chooses the RNG queue until the older RNG requests are scheduled to prevent starvation of RNG applications.} After RNG requests are scheduled, it {prioritizes} the regular read queue again.} 

{\textbf{Equal Priorities.} {If two RNG applications with an RNG request have the same priority level, the scheduler schedules the older RNG request first. If both RNG applications have regular memory requests, the scheduler schedules them by following the baseline scheduling policy. Similarly, in between two non-RNG applications' requests with the same priority level, the scheduler follows the rules of the baseline scheduler. However, if} {one RNG application with an RNG request and one non-RNG} application {with a regular request} have the same priority, \mechanism{} prioritizes the RNG {requests} to minimize the RNG interference. {Section~\ref{priorityresults}} {show{s} that \mechanism{} does not degrade the performance of non-RNG applications or system fairness when {the} RNG queue is prioritized.}}

{\textbf{Starvation Prevention Mechanism. }There can be extreme cases that can lead to starvation regardless of the RNG-aware scheduling rules we explained. An RNG application that requires a high RNG throughput can fill the RNG queue frequently. If this application is prioritized, then the regular read requests can experience long memory stalls. 
Similarly, a high-priority memory-intensive non-RNG application can fill the regular read queue frequently and can cause RNG applications to experience long memory stalls. The RNG-aware scheduling rules might not be sufficient to identify these cases. Therefore, we employ a starvation prevention mechanism.}

{The prevention mechanism works as follows. \mechanism{} observes which memory request queue is deprioritized due to the priority levels of running applications and prevents the scheduler from stalling the deprioritized queue for a long time period. RNG-aware scheduler increments a counter, \emph{the stall time counter}, when it chooses a memory request queue {based on the higher priority level of an application}, which means the other queue is deprioritized and stalled. It then compares the stall time counter to a threshold value, \emph{the stall limit}, and if the stall time counter reaches the stall limit, the scheduler chooses a request from the deprioritized queue. This prevents high-priority applications from starving other types of applications. In every cycle that a request from the deprioritized queue is scheduled or {when} the priority levels {are} changed, the stall time counter is set to 0.}

{When we set the stall limit as 100 memory cycles {in our evaluation}, none of the workloads cause the stall time counter to reach the stall limit. We observe that the RNG-aware scheduler's rules are sufficient to prevent starvation even when a memory-intensive non-RNG application or an RNG application with a high RNG {throughput} requirement is prioritized.}

%% file: sections/6_3_interface.tex
\subsection{Application Interface}
\label{App Interface}
An interface between the software and the DRAM{-based T}RNG mechanism is needed to integrate the mechanism into the system. In Linux systems, this can be done by changing the existing interface of the kernel's random number generator ~\cite{gutterman2006analysis}. The random number generator uses environmental noise from device drivers and gathers bits in an entropy pool. The \textit{getrandom()} system call is used when an application asks for random numbers, which fills a buffer passed to it with a pointer using the random bits in the entropy pool. 

Our proposed interface changes {the} \textit{getrandom()} system call to use \mechanism{}. This can be done by setting memory-mapped configuration status registers or using other existing I/O datapaths based on the target system. When a request is made, \mechanism{} serves the request either from the {random number} buffer {(see Section~\ref{Random Number Buffer})} or by generating random numbers {(see Section~\ref{RNGA Scheduler})}.

{When the random number buffer is not empty, \mechanism{}'s system call does not incur any overhead over the baseline system call. However, when the random number buffer is empty{,} the overhead is related to the difference between the latencies of gathering random bits with the available RNG of the baseline system (e.g., the Linux kernel's built-in RNGs~\cite{gutterman2006analysis}) and DRAM-based TRNGs. Since the first one depends on available devices that can be used as entropy sources, this overhead depends on the target system.}

%% file: sections/6_4_security_analysis.tex
\revision{\section{Security Analysis of \mechanism{}}}

\revision{\textbf{Secure Random Numbers.} \nise{To} use random numbers for security-critical applications, a good TRNG should provide two key properties. First, the TRNG should not leak random numbers to {applications} other than the requesting {one}. \mechanism{} ensures this property by using a random number buffer that can be accessed by applications only {via} a system call. The system call returns random bits only to the application that requested them and discards the used bits from the buffer. Second, the TRNG should provide {a} unique random number to each random number request. \mechanism{} ensures this property by discarding each random number after being served. We conclude that the simple and restrictive interface of \mechanism{} provides secure true random numbers to security-critical applications.}

\revision{\textbf{The Random Number Buffer as a Timing Side-Channel.} Side-channel attacks~\cite{kocher1996timing} are a {class} of attacks that observe and use side-channel information to infer application behaviour or leak sensitive information and confidential data~\cite{kocher1996timing}. Unlike many other shared hardware structures (e.g., caches {~\cite{liu2013experimental,kim2012stealthmem,wang2007new})} that can be exploited to provide timing information about arbitrary data, \mechanism{} can be exploited by an attacker to measure the time it takes to get a random number from \mechanism{}. This timing information can be used to infer {whether or not} the buffer is empty, {and} understand if another application is requesting random numbers. This side channel is difficult to exploit efficiently {due to two} reasons: (1) \mechanism{} continuously fill{s} the buffer with random numbers {asynchronously to the applications} {and this} could cause the buffer to be empty very few times, and {(2)} if two or more applications are using random numbers, it might be challenging to identify which of those applications is emptying the buffer. We conclude that \mechanism{} timing side channels are {likely to be more difficult} to exploit than side-channels from other {more general-purpose} shared hardware structures (e.g., caches).}

\revision{\textbf{The Random Number Buffer as a Covert Channel.} An attacker can use channels that are not designed as communication channels to send data from one process to another. 
The random number buffer can be used as a covert channel~\cite{lampson1973note} under certain circumstances{,} such as when no process {other than} an attacker controlled process is requesting random numbers. The random number buffer provides a covert channel fundamentally similar to existing cache covert channels. Previous work proposes several techniques to use shared caches as covert channels ~\cite{ristenpart2009hey,percival2005cache,maurice2015c5,wu2014whispers,xu2011exploration,liu2015last,hunger2015understanding} and countermeasures to mitigate such attacks ~\cite{kim2012stealthmem,wang2007new,martin2012timewarp,hu1992reducing,liu2016catalyst,wang2016secdcp}.
Countermeasures of cache-based covert channels can be applied to the random number buffer. First, the {random number} buffer {can be partitioned} {across} different threads with {small} performance overhead, {since} {a} small size buffer {is} effective {for many applications} as shown in Section \ref{bufferEval}. Second, the system can give access privilege of the buffer {only} to a small number of applications (possibly to one application at a time). This would hurt the performance of RNG applications that cannot access the random number buffer {at a given time,} but these applications would still benefit from \mechanism{}’s RNG-Aware scheduler design.}

\revision{\textbf{Denial of Service Attacks.} An attacker might {attempt to} occupy the DRAM bandwidth with RNG requests to starve other applications. {Our RNG-aware memory request scheduler prevents applications from starving with a set of rules to provide system fairness (see Section~\ref{RNGA Scheduler}).} In addition to these rules, DoS attacks can be easily mitigated at OS-level by adapting fairness-aware process scheduling policies (e.g., \cite{cfs}) to be aware of RNG requests. } 

\pagebreak

%% file: sections/7_1_methodology.tex
\section{Methodology}
\label{methodology}

To evaluate \mechanism{}'s performance and fairness, we use a modified version of Ramulator \cite{kim2016ramulator, ramulatorgithub}, a cycle-accurate DRAM simulator, that can simulate two state-of-the-art DRAM-based TRNG mechanisms (i.e., D-RaNGe~\cite{drange} and QUAC-TRNG~\cite{olgun2021quactrng}).  {We extend the core model of Ramulator to support random number requests.} We simulate a system with configurations given in Table \ref{configs}. We use the configurations of \mechanism{} in Table \ref{configs} and simulate {the T}RNG as proposed in D-RaNGe~\cite{drange} unless stated otherwise.

\begin{table}[b]
\centering
\caption{Simulated System Configuration}
\label{configs}
\resizebox{\linewidth}{!}{
\begin{tabular}{ll}
\hline
Processor                                                   & \begin{tabular}[c]{@{}l@{}}1-2-4-8-16 cores, 4GHz clock frequency,\\ 3-wide issue, 128-entry instruction window\end{tabular}  \\ \hline
DRAM                                                        & \begin{tabular}[c]{@{}l@{}}DDR3-1600 ~\cite{jedec2012}, 800Mhz bus frequency,\\4 channels, 1 rank/channel,\\8 banks/rank, 64K rows/bank\end{tabular}  \\ \hline
\begin{tabular}[c]{@{}l@{}}Memory Ctrl.\end{tabular} & \begin{tabular}[c]{@{}l@{}}32-entry read/write queues,\\ FR-FCFS~\cite{rixner2000memory,zuravleff1997controller}{~with a column cap of 16~\cite{mutlu2007stall}}\end{tabular}   \\ \hline
DR-STRANGE                                                  & \begin{tabular}[c]{@{}l@{}}32-entry random read queue, RNG-aware\\ scheduler, 256-entry predictor table/channel, \\ 16-entry random number buffer\end{tabular} \\ \hline
\end{tabular}}
\end{table}

{\textbf{Workloads. }We evaluate 43 single-core applications from five benchmark suites: SPEC CPU2006 \cite{spec2006}, TPC \cite{tpcweb}, STREAM \cite{McCalpin2007}, MediaBench \cite{fritts2009media}, and YCSB benchmark {suite}~\cite{ycsb}. 
{Based on the last-level cache misses-per-kilo-instruction (MPKI), we group the applications into three categories: (1) L (low memory-intensity, MPKI $<$ 1), (2) M (medium memory-intensity, 1 $\leq$ MPKI $<$ 10), (3) H (high memory-intensity, MPKI $\geq$ 10). To do so, we obtain}
the MPKI values of the applications by analyzing SimPoint \cite{simpoint} traces (200M instructions) of each application.}

We create synthetic {RNG} benchmarks with {varying} random number request {intensities} to test our designs. The random number request intensity is controlled by the number of instructions the benchmark has in between two {64-bit} random number requests. The {synthetic RNG}
applications read from all banks across all channels, but they are not memory intensive {in terms of} {non-RNG} requests. The least {RNG-intensive} synthetic benchmark requests {random numbers with a throughput of} $640Mb/s${,} which is below the maximum throughput of the current best DRAM TRNG mechanism. The most {RNG}-intensive benchmark {requires} $5Gb/s$ random number throughput. We use the {most} {RNG-}intensive synthetic {RNG} benchmark for all tests unless stated otherwise.

{We create 43 two-core workloads{,} each consisting of one non-RNG and one RNG application. In addition, {for {the} four-core configuration} we create four multi-programmed workload groups{,} each consisting of 10 multi-programmed workloads. Each group has 3 different applications from different memory-intensity categories and one synthetic {RNG} benchmark. For example{,} a workload in LLHS group has two randomly selected single-core applications from low memory-intensity category, one randomly selected single core application from high memory-intensity category, and a synthetic {RNG} benchmark. We also create 30 multi-programmed workloads for 8-core and 16-core configurations consisting of low, medium, and high memory-intensity applications. }

\textbf{Comparison Points.} We compare \mechanism{} to (1) an RNG-oblivious system, and (2) {a Greedy Idle Design,} for performance and fairness results. We compare our RNG-aware Scheduler design {to} FR-FCFS~\cite{rixner2000memory,zuravleff1997controller} with a column cap of 16~\cite{mutlu2007stall} and BLISS ~\cite{subramanian2016bliss,subramanian2014blacklisting} memory schedulers. {We compare (1)~the simple DRAM {idleness} predictor {with {a} low utilization threshold of 4} to (2)~{the} RL-based DRAM idleness predictor (see Section~\ref{predictor}).}

{The Greedy Idle Design is} based on the idea of a buffer filling mechanism with no overhead.\footnote{{The Greedy Idle Design provides an upper bound for performance and fairness results using a greedy algorithm. However, its performance and fairness improvement is limited because the best {dynamic} memory request scheduling order cannot be determined with a greedy approach in polynomial time.}} 
If an idle period reaches the \emph{Period Threshold}, which is 40 cycles in our tests, we assume we fill the buffer with 8 random bits without any overhead. {Greedy Idle Design} has a separate RNG read queue similar to \mechanism{}. To simulate a fair comparison point, we also employ application{-}priority-based RNG-aware scheduling in the {Greedy Idle Design}.  

{\textbf{Metrics.} We measure the performance of the non-RNG and RNG applications running on a dual-core system using normalized execution time of the same number of simulated instructions. For multi-core evaluations, for non-RNG applications we use the weighted speedup metric \cite{snavely2000symbiotic}, which prior work shows is a good measure of job throughput \cite{eyerman2008system}. 
}

For fairness results{,} we use the unfairness metric proposed in {\cite{mutlu2007stall,gabor2006fairness,moscibroda2007memory}}. Unfairness index is calculated as the ratio of the maximum memory-related slowdown experienced by an application in the workload to the minimum memory-related slowdown.  To calculate the memory-related slowdown of an application, we measure the memory stall time when memory is shared normalized to the memory stall time when the application runs alone:

\begin{gather*}
\scriptstyle{MemSlowdown_{i} = \frac {MCPI^{shared}_{i}}{MCPI^{alone}_{i}},\:
Unfairness =\frac{ max_{i}\:MemSlowdown_{i}}{min_{i}MemSlowdown_{i}}}
\end{gather*}

If the unfairness index is equal to one, it means that applications running on the system experience similar slowdowns.
A higher unfairness index shows that the system unfairly prioritizes one application, and thus there is a {large difference} between the slowdowns of the applications.

%% file: sections/8_1_perf_analysis.tex
\section{Results}
\label{analysis}

{We evaluate the performance, fairness, energy efficiency, and area overhead of \mechanism{}.}

\subsection{Impact on Performance}
\label{perfResults}
\sloppy

\subsubsection{{Dual}-core Performance}
\begin{figure}[h]
\includegraphics[width=\linewidth]{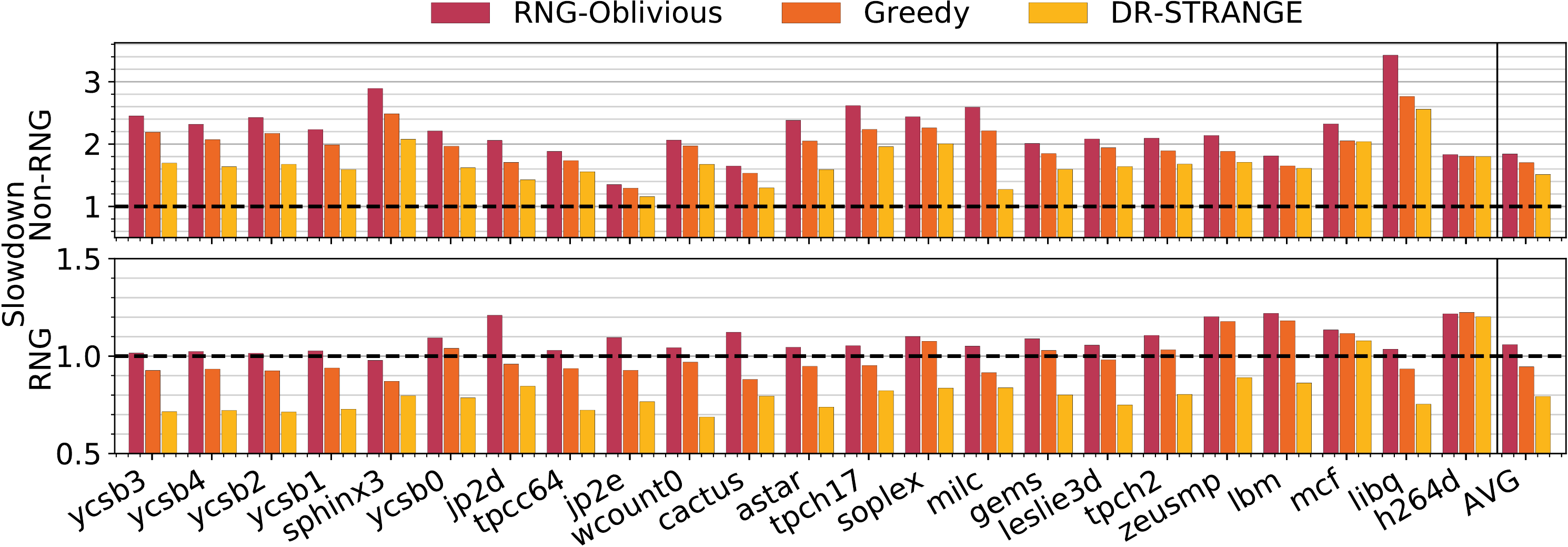}
\caption{{Slowdown} {over single-core execution} of non-RNG (top) and RNG (bottom) applications in {dual}-core workloads.}
\label{fig:final-80}
\end{figure}

{{Figure~\ref{fig:final-80}} compare{s the} performance results of three designs{:} {(1)~{the} RNG-Oblivious baseline, (2)~{the} Greedy Idle Design, and (3)~\mechanism{}}. The figure shows {the} slowdown of non-RNG (top) and RNG (bottom) applications in workloads executed on a dual-core system compared to each application's performance when executed {alone} on a single core. We make {two} observations.}

First, \mechanism{} improves {the} performance of both RNG and non-RNG applications. \mechanism{} reduces the execution time of non-RNG applications by 17.9\% and RNG applications by 25.1\% on average compared to {the baseline} RNG-oblivious system. {\mechanism{} improves {the} average performance of RNG applications by 20.6\% compared to {the} performance of RNG applications when executed {alone} on a single core{,} due to {the} low{er} RNG latency.}

Second, the {greedy} design improves average performance of non-RNG {and RNG} applications by 7.6\% and 10.7\%, {respectively}. \mechanism{} outperforms the {greedy} design in most of the workloads {because} the {greedy} design fill{s} the {random number }buffer only in sufficient{ly long} idle periods. In contrast, \mechanism{} leverages the low DRAM utilization {to fill the random number buffer} with {{its} {low utilization} prediction mechanism (see Section \ref{predictor})}. 

\subsubsection{Multi-core Performance} 
\begin{figure}[ht]
\centering
\includegraphics[width=\linewidth]{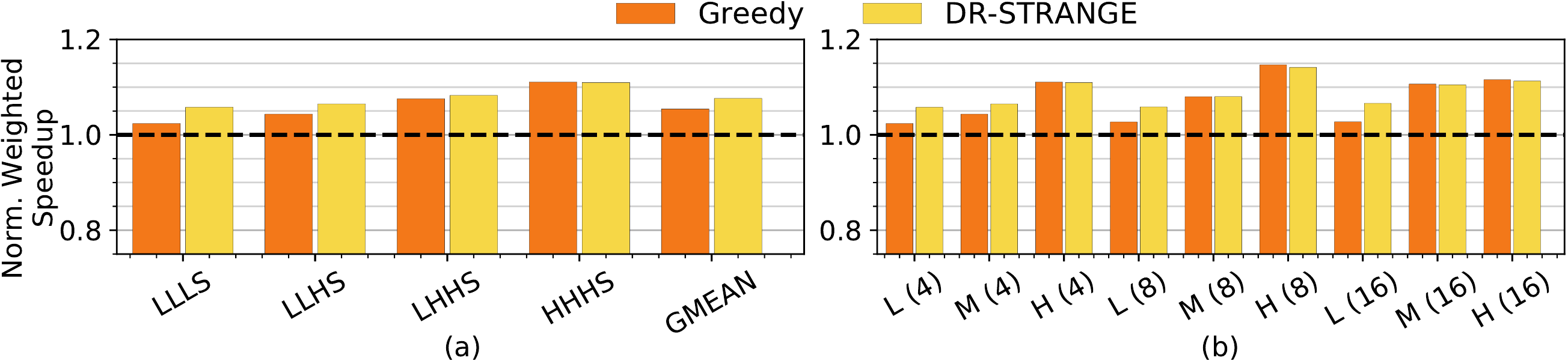}
\caption{{Normalized weighted speedup of non-RNG applications in (a) 4-core workloads, and (b) 4-, 8-, 16-core workloads grouped based on memory-intensity.}}
\label{fig:wl}
\end{figure}

~\\
\textbf{Non-RNG Applications.} Figure \ref{fig:wl} shows {the normalized} weighted speedup of non-RNG applications in four-core workload groups (left) and 4-, 8-, 16-core workload groups (right) normalized to the RNG-oblivious {baseline} design. {W}e make two observations. First, for four-core workloads{,} \mechanism{} {provides} 7.6\% {average performance improvement}. 
{As the} number of high{ly} memory-intensive applications in the workload {increases}, \mechanism{}'s {performance} {improvement} increases. The {greedy} design has 5.4\% {higher} {performance} {improvement over the RNG-oblivious baseline. \mechanism{} outperforms the greedy design because \mechanism{} serves {a larger number of} random number requests from the random number buffer compared to the greedy design.} Second, \mechanism{} {provides} 12.1\%, 8.2\%, and 6.1\% {average} {performance} improvement for workloads consisting of high, medium, and low memory-intensity applications, respectively. {\mechanism{} outperforms greedy design in all workload groups with only two exceptions. For highly memory-intensive workloads consisting of 8 and 16 applications, the greedy design {provides} a slightly higher performance improvement compared to \mechanism{} due to the performance overhead of DRAM utilization mispredictions of \mechanism{}.}

\begin{figure}[ht]
\centering
\includegraphics[width=\linewidth]{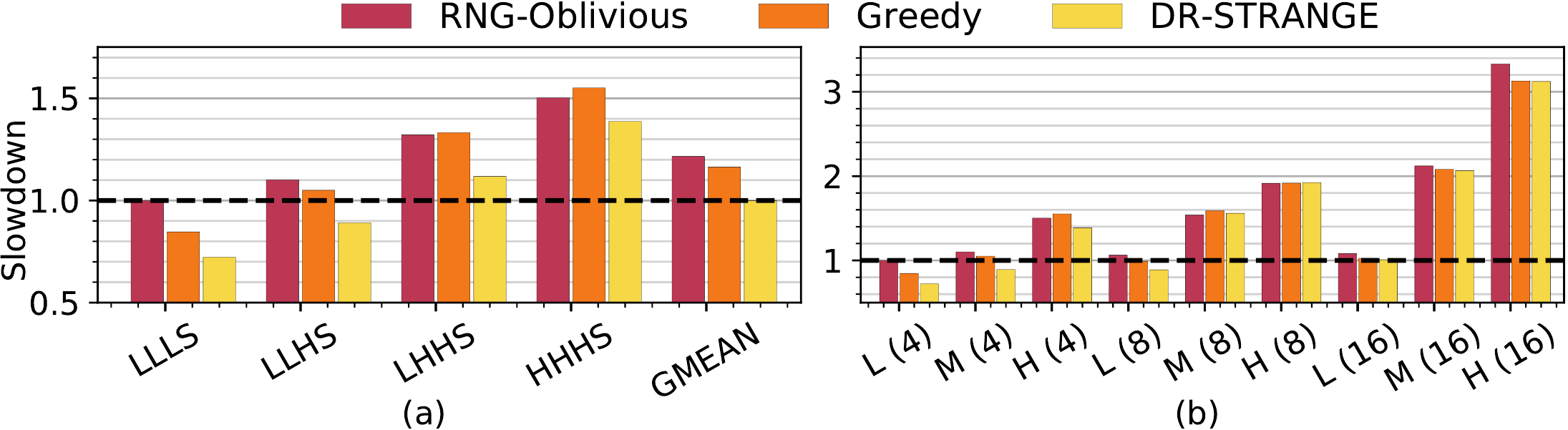}
\caption{{Slowdown of RNG applications in (a) 4-core workloads, and (b) 4-,8-,16-core workloads grouped based on non-RNG {applications'} memory-intensity.}}
\label{fig:wl-s}
\end{figure}

\textbf{RNG Applications.} {Figure \ref{fig:wl-s} compare{s} the performance of RNG applications on (1) {the} RNG-Oblivious baseline, (2) {the} Greedy Idle Design, and (3) \mechanism{}  over the performance of RNG applications when executed on a single-core baseline system. The figure shows {the} {slowdown} of RNG applications in four-core workload groups (left) and 4-, 8-, 16-core workload groups (right). For four-core workload groups{,} \mechanism{} improves average performance by 17.8\%. For 4-, 8-, 16-core workload groups, \mechanism{} improves average performance by 4.5\%, 6.7\% and 16.9\% for high, medium, and low memory-intensity workloads. For all workloads \mechanism{} improves performance {at least as much as} the {greedy idle} design.}

%% file: sections/8_2_fairness_analysis.tex
\subsection{Impact on {System} Fairness}
\label{fairnessResults}
\begin{figure}[h]
\includegraphics[width=\linewidth]{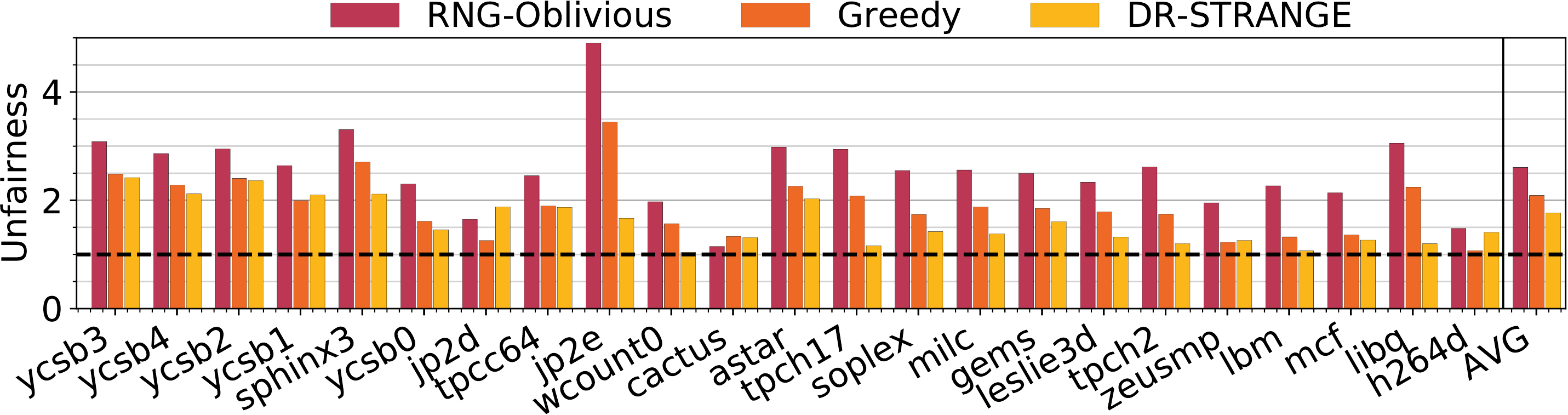}
\caption{{System fairness on a dual-core system.}}
\label{fig:final-fairness}
\end{figure}

{We evaluate system fairness impact of three designs: (1) {the} RNG-Oblivious baseline, (2)~{the} Greedy Idle Design, and (3) \mechanism{}.}
Figure~\ref{fig:final-fairness} plots {system fairness, {which we calculate using the \emph{unfairness index} metric~{\cite{mutlu2007stall,gabor2006fairness,moscibroda2007memory}}, for}} two-core workloads. {We make three observations. First,} \mechanism{} improves average system fairness by 32.1\% compared to the RNG-oblivious {baseline} design. {Second, }\mechanism{} outperforms the \revision{greedy} design {by 15.2\% in terms of} fairness. {Third, some} workloads {that include} non-RNG applications, such as \emph{jp2d} and \emph{cactusADM}, show higher unfairness indices {compared to the greedy design} because \mechanism{} {improves performance of} RNG applications more {than {that of} such non-RNG applications} due to {its} effective {random number} buffering. {We conclude that \mechanism{} outperforms both the RNG-oblivious baseline and the Greedy Idle Design in terms of system fairness by employing an RNG-aware scheduling policy and effectively reducing and controlling the RNG interference. }

%% file: sections/8_3_buffer_analysis.tex
\subsection{Impact of {the} Random Number Buffer}
\label{bufferEval}
{Serving random number requests from the random number buffer reduces the total execution time by reducing the RNG latency. {The impact of the random number buffer depends on the ratio of random number requests served from the random number buffer {over} all random number requests. We call this ratio} \emph{the buffer serve rate}. The buffer serve rate depends on (1) DRAM channel utilization, (2) required TRNG throughput, and (3) the random number buffer size.}
{In this section, we evaluate \mechanism{} with {different} random number buffer sizes maintained with the simple buffering mechanism. }

\begin{figure}[ht]
\centering
\includegraphics[width=\linewidth]{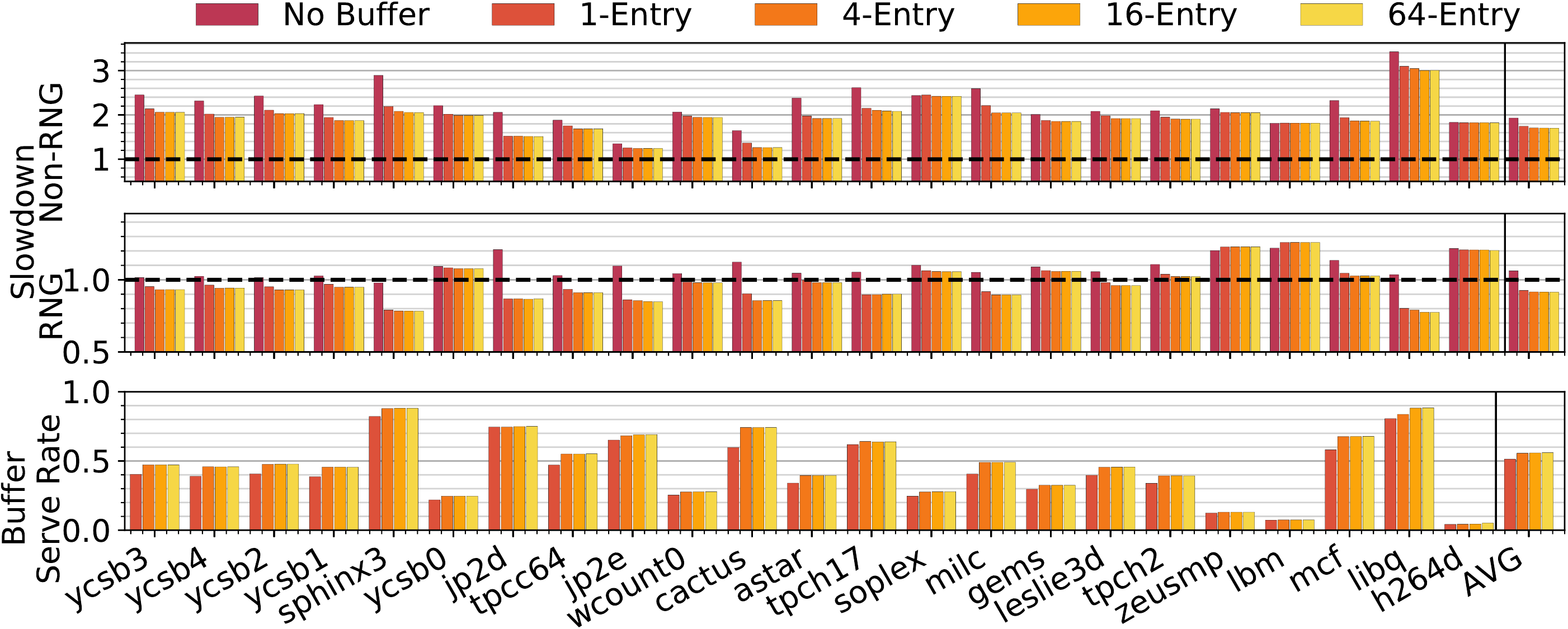}
\caption{{I}mpact of the buffer size on slowdown of non-RNG (top{, lower is better}) and RNG applications (middle{, lower is better}) and buffer serve rate (bottom{, higher is better}).}
\label{fig:buffer}
\end{figure}

{Figure~\ref{fig:buffer} shows the {slowdown} of non-RNG and RNG applications executed on a dual-core system {compared to when each application is} executed {alone} on a single-core system (top and middle) and the buffer serve rate of workloads (bottom). We make two observations.}

{First, {a} 16-entry random number buffer improves the average performance of non-RNG and RNG applications by 11.7\% and 13.8\%, respectively. {T}he performance improvement of non-RNG and RNG applications (top and middle) increases with larger buffer sizes until 16 entries and a significant performance improvement is achieved with a 16-entry random number buffer.} 

{Second, a 16-entry random number buffer achieves an average buffer serve rate of 0.55. Increasing the buffer size improves the buffer serve rate of {only a few} workloads, such as the workloads including \emph{jp2e}, \emph{cactus}, and \emph{libquantum}.}

{Third, RNG applications in some workloads consisting of memory-intensive non-RNG applications, such as \emph{zeusmp} and \emph{lbm}, experience slowdown {due to the low} number of sufficient{ly long} idle periods and {increased RNG interference {as a result of} random number generation for the buffer. These workloads do not benefit {much} from the random number buffer and {have} low buffer serve rates as shown in Figure~\ref{fig:buffer} (bottom).}}

{We conclude that the buffering mechanism improves the average performance of non-RNG and RNG applications by reducing the RNG interference and TRNG latency.}

%% file: sections/8_4_scheduler_analysis.tex
\subsection{Impact of RNG-Aware Scheduling}
\label{SchedulerComparison}

{We evaluate {(1)} \mechanism{}, {(2)} BLISS \cite{subramanian2016bliss,subramanian2014blacklisting}\footnote{We use a value of four for \emph{Blacklisting Threshold} and a value of 10000 cycles for \emph{Clearing Interval} as proposed in \cite{subramanian2016bliss}.}, and {(3)} the RNG-oblivious baseline design with {the} FR-FCFS scheduler~\cite{rixner2000memory,zuravleff1997controller} with a column cap of 16, {called FR-FCFS+Cap}~\cite{mutlu2007stall}. In these tests, we evaluate \mechanism{} designs with no random number buffer.}

\begin{figure}[h]
\centering
\includegraphics[width=\linewidth]{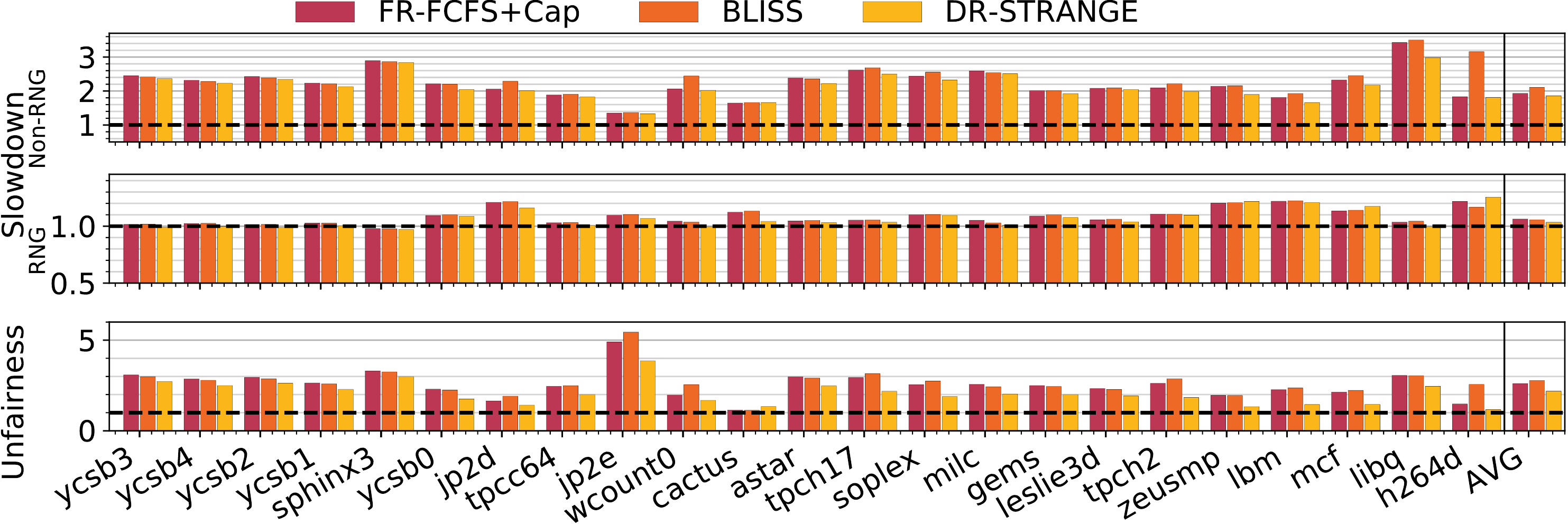}
\caption{Impact of the memory request scheduler on {the} performance of non-RNG (top) and RNG (middle) applications and system fairness (bottom).}
\label{fig:scheduler}
\end{figure}

{Figure \ref{fig:scheduler} shows the performance impact of {the} RNG-Aware scheduler on non-RNG applications (top) and RNG applications (middle) normalized to the performance of {applications executed on a single core}. We make two observations.}

{First, the RNG-Aware scheduler outperforms both FR-FCFS{+Cap} and BLISS mechanisms and improves average fairness by 16.1\%. It improves average performance of non-RNG and RNG applications by 5.6\% and 1.6\%, respectively. The performance of the RNG-aware scheduler is greater than {the} performance of FR-FCFS{+Cap} and BLISS for almost all workloads.} 

{Second, on average, BLISS has a higher unfairness index compared to FR-FCFS{+Cap} and increases average unfairness by 6.6\%.  {Some w}orkloads consisting of memory-intensive non-RNG applications,  such as \emph{jp2e, wcount0, tpch17, soplex, tpch2, lbm, mcf}, and \emph{h264d},  have significantly higher unfairness indices compared to {the} FR-FCFS{+Cap} and the RNG-Aware scheduler{s} because BLISS {frequently} blacklists these high{ly} memory-intensive non-RNG applications and prioritizes RNG applications. Due to {the frequently} blacklisted non-RNG applications, BLISS degrades {average} performance of non-RNG applications over FR-FCFS{+Cap} by 8.9\%. }

\subsection{Impact of Priority-Based Scheduling}
\label{priorityresults}
{To show the impact of priority-based RNG-aware scheduling, we compare three designs: (1) {the} RNG-Oblivious baseline, (2) \mechanism{} with high-priority RNG application{s}, {and} (3) \mechanism{} with high-priority non-RNG application{s}. }

\begin{figure}[h]
\centering
\includegraphics[width=0.85\linewidth]{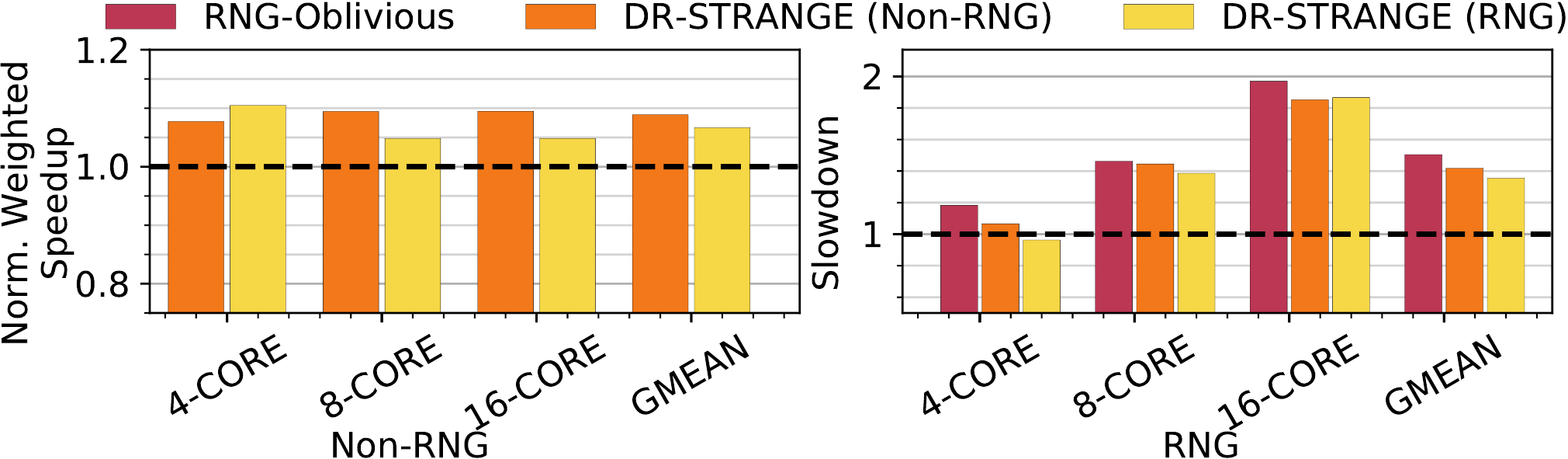}
\caption{Impact of the priority-based RNG-aware scheduler on performance of non-RNG (left) and RNG (right) applications.}
\label{fig:scheduler-wl}
\end{figure}

{Figure \ref{fig:scheduler-wl} shows the normalized weighted speedup of non-RNG applications (left) and the {slowdown} of RNG applications normalized to their performance when executed on a single core (right). We make three observations.}

{First, priority-based RNG-aware scheduling {provides} significant performance improvement when workloads fully utilize the DRAM bandwidth. Both types of applications benefit from RNG-aware scheduling and performance  improvements increase when they have {high priority} levels.}

{Second, Figure \ref{fig:scheduler-wl} (left) shows that prioritizing {the} non-RNG applications using \mechanism{} improves {the} average weighted speedup of {the} non-RNG applications by 8.9\% compared to the RNG-oblivious baseline. Figure \ref{fig:scheduler-wl} (right) {shows that} prioritizing {the} RNG applications using \mechanism{} improves average performance of {the} RNG applications by 9.9\% compared to the RNG-oblivious {baseline}.}

{Third, prioritizing RNG applications using \mechanism{} improves {the} average performance of both non-RNG and RNG applications in 4-core workloads. Some workloads consisting of low and medium memory-intensity non-RNG applications switch between RNG and regular {memory request} queues frequently when non-RNG applications are prioritized due to low memory-intensity applications creating requests at lower rates. These workloads benefit from lower RNG interference when RNG applications are prioritized using \mechanism{}.}

%% file: sections/8_5_pred_analysis.tex
\subsection{Impact of the DRAM {Idleness} Predictor}
\label{pred}

\begin{figure}[ht]
\centering
\includegraphics[width=\linewidth]{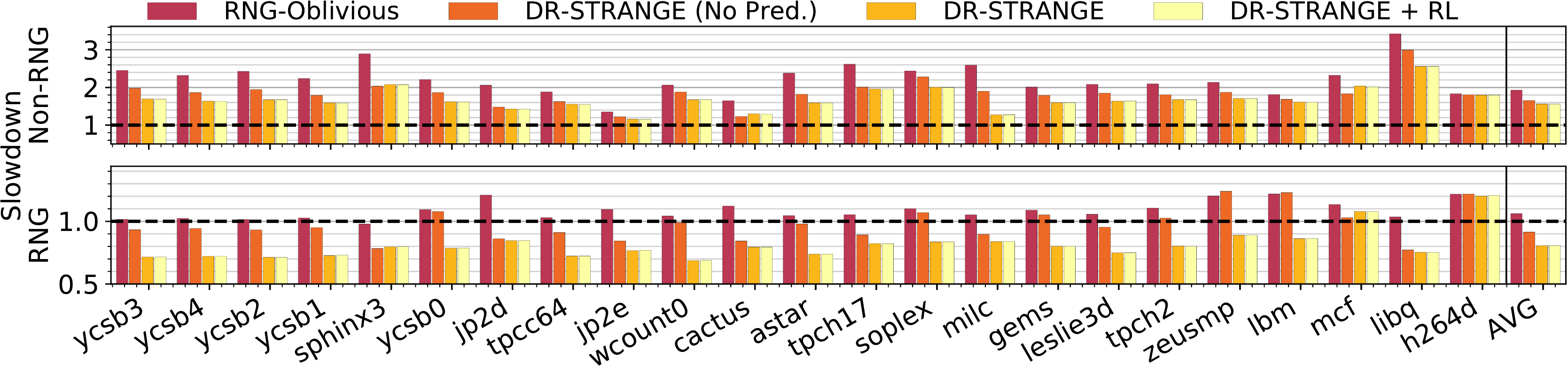}
\caption{Impact of {the} DRAM {Idleness} Predictor on non-RNG (top) and RNG (bottom) applications' performance. }
\label{fig:pred-perf}
\end{figure}

{Figure \ref{fig:pred-perf} evaluates the performance of four designs: (1) {the} RNG-oblivious baseline, (2) \mechanism{} {with no DRAM idleness predictor}, (3) \mechanism{} with the simple DRAM {idleness} predictor, and (4) \mechanism{} with {the} RL-based DRAM idleness predictor. Figure \ref{fig:pred-perf} plots the slowdown of non-RNG (top) and RNG applications (bottom). {We make three observations.}}
{First,} \mechanism{} with {the simple} DRAM {idleness} predictor 
{improves average performance of non-RNG and RNG applications by 17.9\% and 25.1\%, respectively, compared to {the} RNG-oblivious {baseline}.}
{Second, {the simple} DRAM {idleness} predictor improves \mechanism{}'s performance by 12.4\% and 13.8\% for non-RNG and RNG applications, respectively.}
{Third,} \mechanism{} with {the} simple predictor achieves similar performance improvement {at} 
{lower hardware complexity}
compared to \mechanism{} with {the} RL{-based idleness predictor,} which improves {non-RNG and RNG applications' average performance} by 19.3\% and 23.9\%, respectively.

\begin{figure}[h]
\centering
\includegraphics[width=\linewidth]{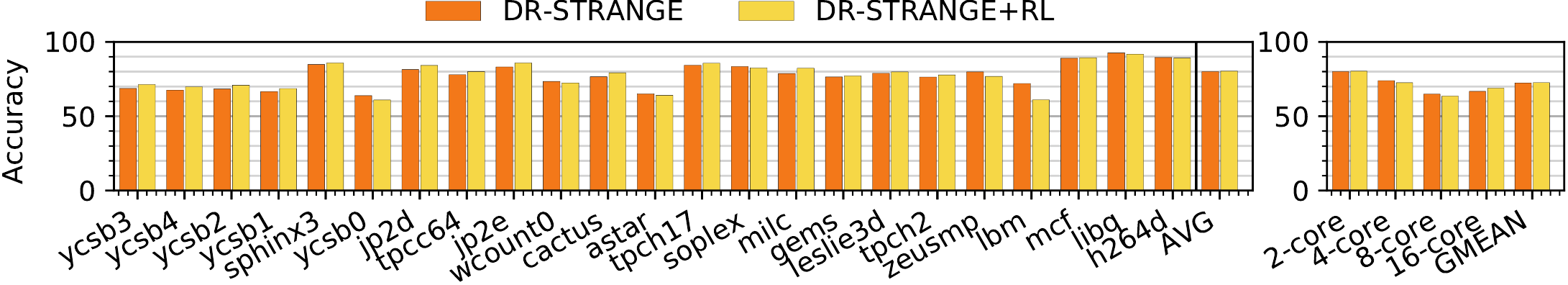}
\caption{DRAM {idleness} predictor accuracy in two-core (left) and multicore workloads (right).}
\label{fig:pred-accuracy}
\end{figure}

Figure \ref{fig:pred-accuracy} shows the {predictor} accuracy of {the} 2-core workloads (left) and 2-, 4-, 8-, 16-core workloads (right). {We make the following observations. First,} with two-core workloads{,} {the simple DRAM idleness} predictor and {the} RL-based idleness predictor have {an accuracy of} 80.0\% and 80.3\%{, respectively. Second,} the {simple} DRAM {idleness} predictor's accuracy is {slightly} higher than the RL-based predictor's accuracy when the workload's memory-intensity is high and the number of {long} idle periods is low. {In} workloads {containing} low memory-intensity applications, the {simple DRAM idleness} predictor mispredicts {long} idle periods to be short and has {a} higher false {negative} rate. {Third,} {in 4-, 8-, and 16-core workloads}{,} both predictors have lower accuracy {due to less idleness and more complex interference patterns}.

\subsubsection{{Impact of the Low Utilization Prediction}}
\begin{figure}[ht]
\centering
\includegraphics[width=\linewidth]{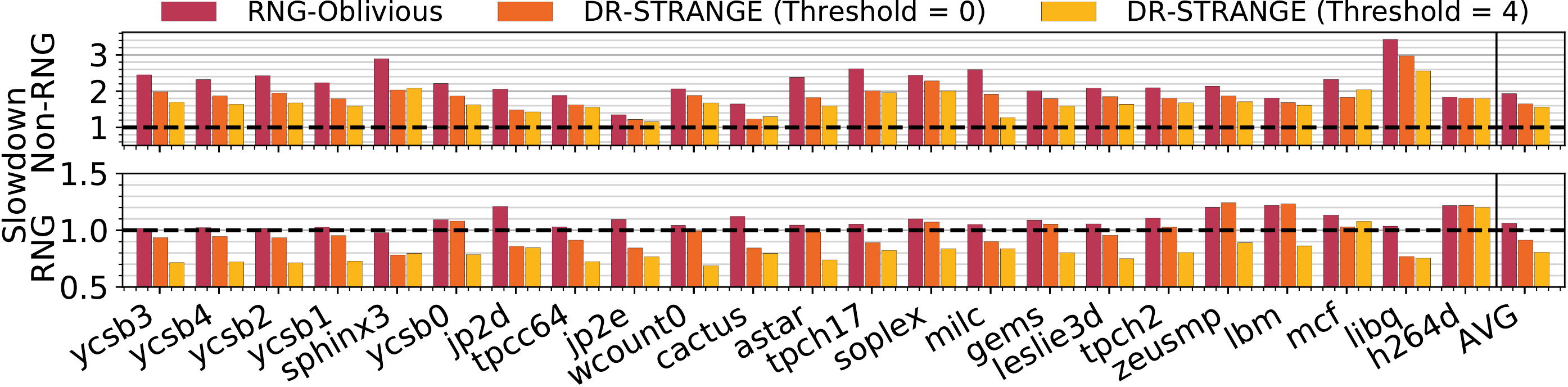}
\caption{{Impact of the low utilization prediction on non-RNG (top) and RNG (bottom) applications' performance.}}
\label{fig:pred-threshold}
\end{figure}

{We evaluate three designs: (1)~the RNG-oblivious baseline, (2)~the simple DRAM idleness predictor without low utilization prediction, and (3)~the simple DRAM idleness predictor with low utilization prediction using the threshold value of 4. Figure \ref{fig:pred-threshold} shows the performance impact of \mechanism{} with and without {the} low utilization prediction mechanism on non-RNG (top) and RNG (bottom) applications. We conclude that the simple DRAM idleness predictor with {a} low utilization threshold of 4 improves the average performance of non-RNG and RNG applications by 5.5\% and 11.7\%, respectively, compared to the DRAM idleness predictor without low utilization prediction.   
}

%% file: sections/8_6_quac_analysis.tex
\subsection{Experiments {U}sing QUAC-TRNG}
\label{quacResults}

{To show {that} \mechanism{} is compatible with different DRAM-based TRNGs, we evaluate \mechanism{} with QUAC-TRNG~\cite{olgun2021quactrng}. QUAC-TRNG {provides} a higher {RNG} throughput compared to D-RaNGe~\cite{drange}. However, it also has {higher latency for 64-bit random number generation}. We compare \mechanism{}'s impact on performance and system fairness compared to the RNG-oblivious baseline when both {systems} use QUAC-TRNG to generate the random numbers.}

\begin{figure}[ht]
\centering
\includegraphics[width=\linewidth]{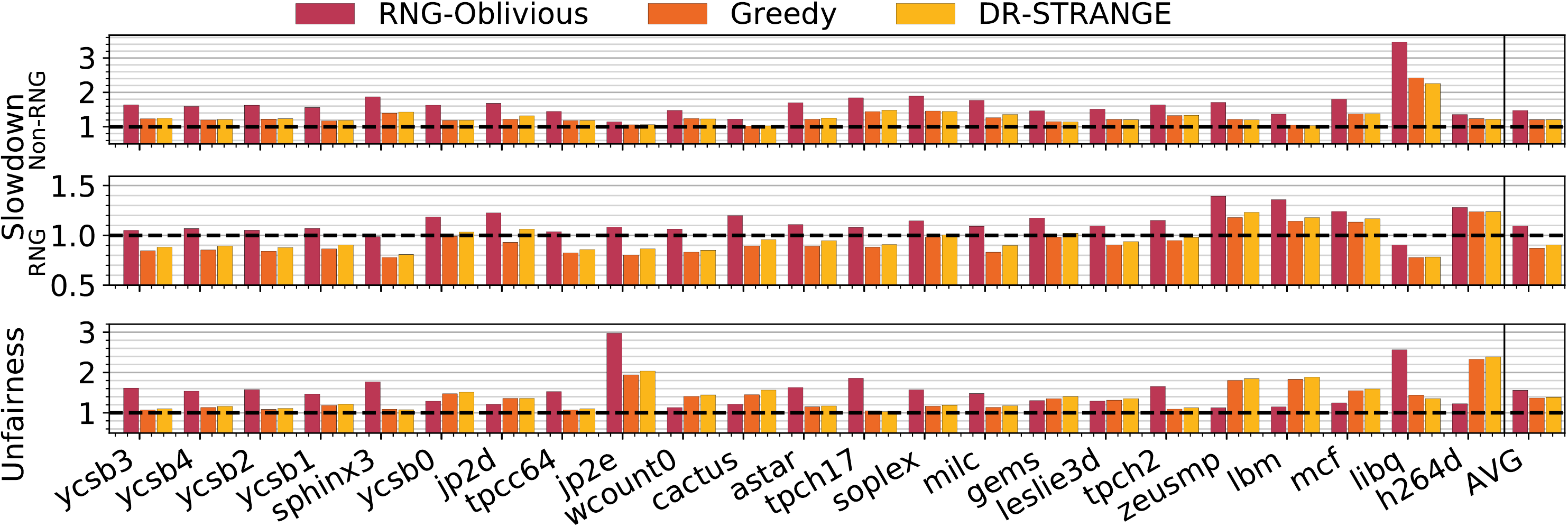}
\caption{{P}erformance and fairness {in} dual-core workloads consisting of RNG and non-RNG applications {in a system that uses QUAC-TRNG~\cite{olgun2021quactrng}}. }
\label{fig:quac}
\end{figure}

{Figure \ref{fig:quac} plots {the} performance of non-RNG and RNG applications (top and middle) and {system} fairness (bottom) compared to the RNG-Oblivious baseline. We make three observations. 
First, \mechanism{} improves average performance of non-RNG and RNG applications by 18.2\% and 17.2\%, respectively. 
Second, \mechanism{} improves average system fairness by 10.9\%. Some workloads with high memory-intensity (\emph{zeusmp, lbm, mcf, h264d}) suffer from higher unfairness indices because \mechanism{} improves performance of non-RNG applications more than the performance of RNG applications.}

{We {conclude} that \mechanism{} improves both performance and system fairness {regardless of {the implemented} DRAM TRNG.}
\mechanism{} can {also} {potentially leverage hybrid {DRAM} TRNGs} that can generate random numbers using (1) {a} low-latency DRAM-based TRNG to fill the random number buffer, and (2) {a} high-throughput DRAM-based TRNG to generate random numbers after receiving {a} random number request when the random number buffer is empty. We leave the evaluation of such design to future work.}

%% file: sections/8_7_other_analysis.tex
\subsection{Results of Low-Intensity Applications}

{We evaluate \mechanism{} with workloads consisting of (1) one low-intensity RNG application (i.e., $640\ Mb/s$ {RNG throughput}), and (2) one non-RNG application. We observe {that} \mechanism{} improves the average performance of {the} RNG and non-RNG applications by 3.2\% and 4.6\% respectively. Due to the low RNG interference caused by the low required TRNG throughput, \mechanism{} does not significantly improve system fairness over the baseline.
}

\subsection{Area and Energy Consumption Analysis}
\label{sec:area}
To evaluate \mechanism{}'s energy consumption, we use DRAMPower \cite{drampower} with {Ramulator's} output traces.
{We observe that \mechanism{} reduces energy consumption and total memory cycles by 21\% and 15.8\%, respectively, compared to {the} RNG-oblivious {baseline} system.}

\revision{We use CACTI~\cite{muralimanohar2009cacti6} at $22nm$ process technology {node} to model {the} DR-STRANGE configuration {shown} in {Table}~\ref{configs}{,} including the random number buffer, RNG request queue, and the DRAM {idleness} predictor. We find that \mechanism{} incurs minor area overhead: 0.0022$mm^2$ (0.00048\% of an Intel Cascade Lake CPU Core~\cite{wikichipcascade}). {These evaluations include the simple DRAM {idleness} predictor that consists of 256 entries and {a} 2-bit saturating counter in each entry, with a total of $0.0625$ KB area overhead. If we use the RL-based predictor, \mechanism{} has an area cost of 0.012$mm^2$ (0.0033\% of an Intel Cascade Lake CPU Core~\cite{wikichipcascade}) and the RL agent requires $8 KB$ {storage} assuming 4-byte Q-values and 10-bit state values.}}

%% file: sections/9_relatedwork.tex
\section{{Other} Related Work}

To our knowledge, \mechanism{} is the first work {to} propose an end-to-end system design for DRAM TRNGs that (1) reduces the interference between RNG and non-RNG applications in the memory controller, (2) improves system fairness, and (3) reduces high TRNG latency. We have already discuss{ed closely} related work on memory request schedulers (Section~\ref{SchedulerComparison}) and DRAM-based TRNG mechanisms and compared \mechanism{} to prior proposals (Sections \ref{fairnessResults}, \ref{perfResults}, and \ref{quacResults}). In this section, we discuss other related works.  

\textbf{Memory Request Scheduling.} Previously proposed memory request scheduler designs~{~\cite{mutlu2009parallelism,kim2010atlas,kim2010thread,ghose2013criticality,nesbit2006fair,mutlu2007stall,ebrahimi2011parallel,ausavarungnirun2012staged,usui2016dash,zhao2014firm,subramanian2014blacklisting,subramanian2016bliss,mutlu2007stall,mutlu2008distributed,hur2004adaptive}} aim to reduce inter-application interference to improve the system performance and fairness. {These proposals are RNG-oblivious. We already compare \mechanism{} to BLISS~\cite{subramanian2014blacklisting,subramanian2016bliss} and FR-FCFS~\cite{rixner2000memory,zuravleff1997controller} with a column cap~\cite{mutlu2007stall} in Section~\ref{SchedulerComparison}.}

\textbf{Low-Throughput DRAM-based TRNGs.}
Prior work propose{s} DRAM-based TRNGs that generate random numbers using different entropy sources, such as DRAM start-up values~\cite{eckert2017drng}, and retention failures~\cite{tehranipoor2016robust,keller2014dynamic,sutar2018d,hashemian2015robust}. These DRAM-based TRNGs are limited in the throughput they can provide and cannot be used in a streaming manner. {As such, they are less practical than the high-throughput DRAM-based TRNGs we evaluate~\cite{drange,olgun2021quactrng}.}

{\textbf{DRAM Idleness Predictors.}
Prior work proposes DRAM idleness predictors that predict the idle period lengths to reduce DRAM energy consumption~\cite{thomas2012pred} and efficiently schedule last level cache writebacks~\cite{wang2012rankidle}.} 
{Compared to these techniques, \mechanism{}'s idleness predictor can be implemented using simple hardware at low cost (Section~\ref{sec:area}) and requires no modifications to the interface between the processor and the memory controller.}

%% file: sections/10_conclusion.tex
\section{Conclusion}
We propose \mechanism{}, {the first}  end-to-end system design for DRAM true random number generators {(TRNGs)} that reduces the RNG interference in the memory controller, provides {system} fairness among different types of applications, and successfully hides the {latency of DRAM-based TRNGs}.  
Our system design consists of three main parts: (1) a random number buffering mechanism combined with {a} DRAM {idleness} predictor, (2) {an} RNG-aware {memory request} scheduler, and (3) {an} application interface. Our evaluations show that \mechanism{} improves the performance of RNG and non-RNG applications by 17.9\% and 25.1\% respectively, and the overall system fairness by 32.1\% while reducing {average} {system }energy consumption {by 21\%.} {We conclude that with an end-to-end system design, DRAM-based TRNGs can be {seamlessly} integrated into today's systems with low overheads and provide true random numbers \nisf{at high throughput and low latency}.}

%% file: sections/11_appendix.tex
\section{Appendix}
\subsection{Analysis of RNG Applications with Higher RNG Throughput Requirements}
\begin{figure}[h]
\centering
\includegraphics[width=\linewidth]{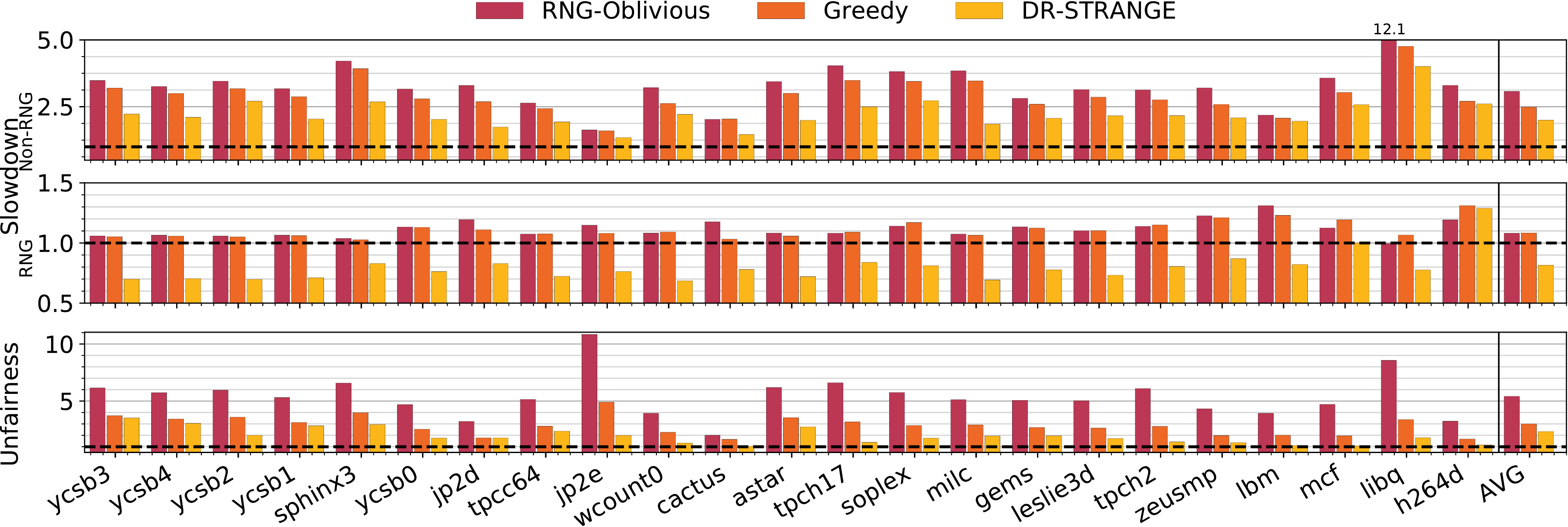}
\caption{{P}erformance and fairness {in} dual-core workloads consisting of non-RNG applications and RNG applications that require $10\:Gb/s$ RNG throughput.}
\label{fig:final-extended}
\end{figure}

Figure~\ref{fig:final-extended} compares the performance and fairness results of three designs: (1) the RNG-oblivious baseline, (2) the Greedy Idle Design, and (3) \mechanism{}. The figure shows the slowdown of non-RNG applications (top) and RNG applications that require $10\:Gb/s$ RNG throughput (middle) executed on a dual-core system compared to each application's performance when executed alone on a single core. Figure~\ref{fig:final-extended} (bottom) plots system fairness, calculated using the unfairness index metric~{\cite{mutlu2007stall,gabor2006fairness,moscibroda2007memory}}. We make two observations. First, \mechanism{} improves the average performance of non-RNG and RNG applications by 34.9\% and 24.5\%, respectively. Second, \mechanism{} improves the average system fairness by 56.9\%, compared to the RNG-oblivious baseline design. 

\subsection{Multicore Workloads}
Table~\ref{table:motivworkloads} shows the dual-core workloads that we use in Section~\ref{Motiv}.
We create 172 workloads, each workload consisting of one non-RNG and one RNG application. We use 43 non-RNG applications across different benchmark suites and create 4 synthetic RNG benchmarks with required RNG throughputs ranging from $640\:Mb/s$ to $5120\:Mb/s$. 

\begin{table} [h]
\centering
\caption{RNG applications}
\resizebox{\linewidth}{!}{%
\begin{tabular}{|l||l|l|} 
\hline
\textbf{Multicore Workloads} & \textbf{Non-RNG Applications} & \textbf{RNG Applications}                                                                                                                                                                                                                                                                          \\ 
\hline
2-core                       & 43 applications               & \begin{tabular}[c]{@{}l@{}}x1 RNG application with 640 Mb/s RNG throughput requirement\\x1 RNG application with 1280 Mb/s RNG throughput requirement\\x1 RNG application with 2560 Mb/s RNG throughput requirement\\x1 RNG application with 5120 Mb/s RNG throughput requirement\\\end{tabular}  \\
\hline
\end{tabular}
}
\label{table:motivworkloads}
\end{table}

\raggedbottom

Table~\ref{table:evaluationworkloads} shows the multicore workloads that we use to evaluate \mechanism{} in Section~\ref{analysis}. We create 186 workloads for 2-, 4-, 8-, and 16-core systems. {We create 43 two-core workloads{,} each consisting of one non-RNG and one RNG application. In addition, {for {the} four-core configuration} we create four workload groups{,} each consisting of 10 multi-programmed workloads. Each group has 3 different applications from different memory-intensity categories and one synthetic {RNG} benchmark. We also create 30 multi-programmed workloads for 8-core and 16-core configurations consisting of low, medium, and high memory-intensity applications. }

\begin{table}[h]
\centering
\caption{Multicore workloads in Section~\ref{analysis}}
\label{table:evaluationworkloads}
\resizebox{\linewidth}{!}{%
\begin{tabular}{|l||l|l|} 
\hline
\begin{tabular}[c]{@{}l@{}}\\\textbf{Number of Cores}\end{tabular} & \textbf{Non-RNG Applications/Workloads}                                                                                                  & \textbf{RNG Applications}                                                                                                                                                \\ 
\hline
2-core                                                             & 43 non-RNG applications                                                                                                                  & \begin{tabular}[c]{@{}l@{}}x1 RNG application with $5120\:Mb/s$ RNG throughput requirement\\x1 RNG application with $640\:Mb/s$ RNG throughput requirement\end{tabular}  \\ 
\hline
4-core                                                             & \begin{tabular}[c]{@{}l@{}}40 workloads consisting of non-RNG applications\\(4 memory-intensity groups, 10 workloads each.)\end{tabular} & x1 RNG application with $5120\:Mb/s$ RNG throughput requirement                                                                                                          \\ 
\hline
8-core                                                             & \begin{tabular}[c]{@{}l@{}}30 workloads consisting of non-RNG applications\\(3 memory-intensity groups, 10 workloads each.)\end{tabular} & x1 RNG application with $5120\:Mb/s$ RNG throughput requirement                                                                                                          \\ 
\hline
16-core                                                            & \begin{tabular}[c]{@{}l@{}}30 workloads consisting of non-RNG applications\\(3 memory-intensity groups, 10 workloads each.)\end{tabular} & x1 RNG application with $5120\:Mb/s$ RNG throughput requirement                                                                                                          \\
\hline
\end{tabular}
}
\label{table:evaluationworkloads}
\end{table}

\subsection{Idle DRAM Period Length Analysis of Multicore Workloads}

\begin{figure}[h]
\centering
\includegraphics[width=0.60\linewidth]{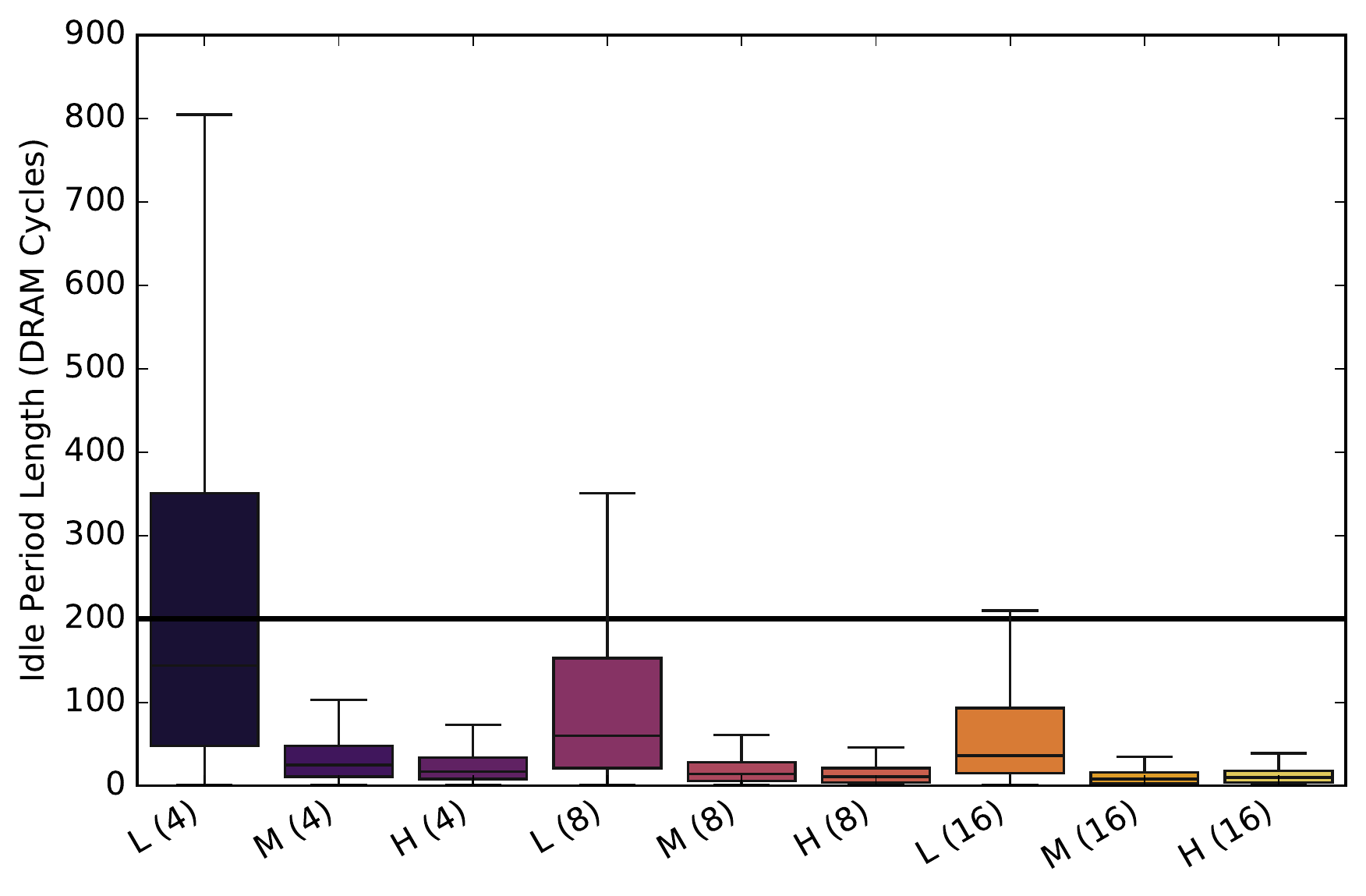}
\caption{{Distribution of} DRAM Idle Period Lengths}
\label{fig:idletime-wl}
\end{figure}

{Figure \ref{fig:idletime-wl} plots the distribution of the idle DRAM period lengths of 4-, 8-, 16-core workloads consisting of non-RNG single-core applications. Based on the memory intensity and the number of applications in each workload, we group these workloads into different \emph{workload groups} and label each with (1) the memory intensity of applications in the group (\emph{L}, \emph{M}, and \emph{H} for low, medium, and high memory intensity respectively), and (2) the number of applications in each workload.  {The distribution is represented as a box-and-whiskers plot where the y-axis {shows} the length of the observed idle DRAM periods in DRAM cycles. The {straight} horizontal line represents the time needed to generate a 64-bit random number, which is} {198 memory cycles ($990\:ns$) on average using D-RaNGe~\cite{drange}} in our test setup
{that {we describe} in {Section} \ref{methodology}}. Each box represents the interquartile range of the observed idle DRAM periods for one workload group and the middle line shows the median. We make two observations. First, similar to our observations regarding the distribution of idle periods for single-core applications (Section~\ref{Random Number Buffer}), a significant portion (84.3\%) of the idle periods do not pass the threshold value of 198 cycles. Second, idle DRAM period lengths decrease with increasing number of applications in the workload and with increasing memory intensity. We conclude that, for many multicore workloads, the majority of the idle periods is not sufficient to generate 64-bit random numbers. }

